\documentclass[12pt]{article}

\usepackage[top=1in,bottom=1in, left=1in, right=1in]{geometry}
\usepackage{bbm}
\usepackage{mathtools}
\usepackage{enumitem}
\usepackage{amssymb}
\usepackage{graphicx}
\usepackage{amsmath}
\usepackage{xcolor}
\usepackage{relsize}
\usepackage{breqn}
\usepackage{bbm}
\usepackage{amsmath,amsthm,amssymb,amsopn,amsfonts,pdfpages,algorithmic,algorithm,dsfont}
\usepackage{graphics}
\usepackage{graphicx}
\usepackage{bbm}
\usepackage{caption,subcaption,url}
\usepackage{stackengine}
\usepackage[colorlinks=true,linkcolor=cyan,citecolor=blue]{hyperref}
\usepackage{authblk}
\usepackage{multirow}
\usepackage{acronym}

\DeclareMathOperator*{\argmax}{arg\,max}
\DeclareMathOperator*{\argmin}{arg\,min}

\newtheorem{theorem}{Theorem}

\newtheorem{defn}{Definition}[section]
\newtheorem{rem}{Remark}
\newcommand{\Ex}{\mathbb{E}}
\newcommand{\p}{\mathbb{P}}
\newcommand{\R}{\mathbb{R}}
\newcommand{\bh}{ {\bf H}}
\newcommand{\bg}{ {\bf G}}
\newcommand{\bde}{\begin{defn}}
\newcommand{\ede}{\end{defn}}
\DeclareMathOperator{\trace}{trace}

\newcommand{\mm}{\mathcal{M}^c_m}
\newcommand{\gn}{\mathcal{G}_n}
\newcommand{\gm}{\mathcal{G}_m}
\newcommand{\mn}{\mathcal{M}^c_n}
\newcommand{\ha}{\widehat{A}}
\newcommand{\hb}{\widehat{B}}
\newcommand{\hc}{\widehat{C}}
\newcommand{\hd}{\widehat{D}}

\newcommand{\tba}{\widetilde{{\bf A}}}
\newcommand{\tbb}{\widetilde{{\bf B}}}
\newcommand{\hba}{\widehat{{\bf A}}}
\newcommand{\hbb}{\widehat{{\bf B}}}

\providecommand{\keywords}[1]
{
  \small	
  \textbf{\textit{Keywords---}} #1
}
\begin{document}
\title{Multiplex graph matching matched filters}

\author[$1$]{Konstantinos Pantazis}
\author[$2$]{Daniel L. Sussman}
\author[$3$]{Youngser Park}
\author[$1$]{Zhirui Li}
\author[$4$]{Carey E. Priebe}
\author[$1$]{Vince~Lyzinski}

\affil[$1$]{\small Department of Mathematics, University of Maryland, College Park
   }
\affil[$2$]{\small Department of Mathematics and Statistics, Boston University}
\affil[$3$]{\small Center for Imaging Sciences, Johns Hopkins University}
\affil[$4$]{\small Department of Applied Mathematics and Statistics, Johns Hopkins University}
\maketitle

\begin{abstract}
We consider the problem of detecting a noisy induced multiplex template network in a larger multiplex background network.
Our approach, which extends the framework of \cite{sussman2018matched} to the multiplex setting,  leverages a multiplex analogue of the classical graph matching problem to use the template as a matched filter for efficiently searching the background for candidate template matches. 
The effectiveness of our approach is demonstrated both theoretically and empirically, with particular attention paid to the potential benefits of considering multiple channels.
\end{abstract}
\keywords{Multiplex graphs, graph matching, correlation network models, matched filters}
\section{Introduction and Background}
Multilayer and multiplex networks have proven to be useful models for capturing complex relational data where multiple types of relations are potentially present between vertices in the network \cite{boccaletti2014structure,kivela2014multilayer}. For example, in connectomes (i.e., brain graphs) different edge modalities can represent different synapse types between neurons  \cite{white1986structure}; in social networks different edge modalities can capture relationships in different social network platforms \cite{goga2015reliability}; in scholarly networks different edge modalities can capture co-authorship across multiple classification categories \cite{ng2011multirank}. Moreover, in many applications leveraging the signal across the different layers of the network can lead to better, more robust performance than working within any single network modality \cite{mucha2010community,kivela2014multilayer,jointLi}.

The inference task we consider here is the problem of detecting (possibly multiple copies of) a noisy induced subgraph in a multiplex background network (see Definition \ref{def:multiplex} for the definition of multiplex networks we consider herein).
Succinctly, given a multiplex template ${\bf A}$ with $m$ vertices, we seek to find the ``best fitting'' subgraph(s) in a larger multiplex background network ${\bf B}$ (see Section \ref{sec:mgmmf} for detail) with $n\gg m$ vertices. This problem is a generalization of the NP-complete \cite{read1977graph} multiplex subgraph isomorphism problem (see \cite{kivela2017isomorphisms} for a definition of multiplex isomorphism), accounting for the reality that relatively large, complex subgraph templates may only errorfully occur in the larger background network. 
These errors may be due to missing edges/vertices in the template or background, and arise in a variety of real data settings  \cite{priebe2015statistical}. 
The subgraph isomorphism problem---given a template $A$, determine if an isomorphic copy of $A$ exists in a larger network $B$ and find the isomorphic copy (or copies) if it exists---has been the subject of voluminous research in the monoplex (i.e., single layer) setting, with approaches based on efficient tree search \cite{efficienttreesearch}, color coding \cite{colorcoding1,colorcoding2}, graph homomorphisms \cite{graphhomobased}, rule-based/filter-based matchings  \cite{feasibilityrules,moorman2018filtering}, among others; for a survey of the literature circa 2012, see \cite{sgsurvey}. 
In contrast, the problem of multilayer homomorphic/isomorphic subgraph detection is still in its relative infancy, with comparatively fewer existing methods in the literature; see, for example, \cite{yang2016mining,takes2017detecting,moorman2018filtering}. 
\vspace{2mm}

\noindent{\bf Notation:} 
The following notation will be used throughout.
For an integer $n>0$, we will define $[n]:=\{1,2,\ldots,n\}$,
$J_n$ to be the $n\times n$ hollow matrix with all off-diagonal entries identically set to $1$,
${\bf 0}_{n}$ to be the $n\times n$ matrix with all entries identically set to $0$.

\subsection{(Multiplex) graph matching}
The above noisy induced subgraph detection problem depends greatly on the definition of ``best fitting'' employed.
Our approach, generalizing \cite{sussman2018matched} to the multiplex setting, will employ the multiplex template ${\bf H}$ to search the multiplex background graph ${\bf G}$ for possible matches, with goodness of fit measured via a multiplex formulation of the classical graph matching problem (see \cite{ConteReview,foggia2014graph,gmrev,yan2016short} for excellent reviews of the voluminous graph matching literature).
In the monoplex setting, the simplest formulation of the graph matching problem (GMP) can be stated as follows:
Given two $n$-vertex, undirected graphs with respective (weighted) adjacency matrices $A$ and $B$, find a permutation matrix $P\in \Pi_{n}=\{n\times n$ permutation matrices\} in 
\begin{align*}
\label{eq:gmp}
\argmin_{P\in\Pi_n}\,\|AP-PB\|_{F}\,=\argmax_{P\in \Pi_n}\,\,\trace(APB^{T}P^{T}).
\end{align*}
Before lifting the graph matching problem to the multiplex setting, we first need to define precisely what we mean by a multiplex graph.

\subsubsection{Multiplex networks}

The above formulation of both the GMP requires both graphs to identically have $n$ vertices, though there are myriad ways of adapting the GMP to graphs of different orders (see, for example, Appendix F of \cite{bento2018family}). 
In the multiplex subgraph matching problem at the core of this paper, we view the template ${\bf A}$ as being equal or lower order than the background ${\bf B}$. Moreover, our definition of multiplex networks, ideally, would allow for differing graph orders across the multiplex layers within a single graph. To allow for these expected data nuances in the multiplex setting, we consider the following multiplex graph model; see \cite{kivela2014multilayer} for a thorough overview of this and other multiplex network formulations. 
\begin{defn}
\label{def:multiplex}
The $c$-tuple ${\bf G}=(G_1,G_2,\dots,G_c)$ is an $n$-vertex multiplex network if for each $i=1,2,\dots,c,$ we have that $G_i\in\mathcal{G}_{n_i}=\{n_i$-vertex labeled graphs$\}$, and the vertex sets $(V_i=V(G_i))_{i=1}^{c}$ further satisfy the following:
\begin{enumerate}[label=\roman*.]
    \item For each $i\in [c]$, we have that $V(G_i)\subseteq [n]$;
    \item $\bigcap\limits_{i=1}^{c}V(G_i)\neq \emptyset$ and $\bigcup\limits_{i=1}^{c}V(G_i)=[n]$;
    \item The layers are a priori node aligned; i.e., vertices sharing the same label across layers correspond to the same entity in the network.
\end{enumerate}
\end{defn}
\noindent Note that each vertex $v\in[n]$ need not appear in each channel $i\in[c]$, however, we do require that at least one vertex appears simultaneously in all channels.
We will denote the set of $c$-layer, $n$-vertex multiplex networks via $\mathcal{M}_n^{c}$. 

\subsubsection{Multiplex GMP}
\label{sec:pad}

To lift the monoplex GMP to the general multiplex definition presented above, we consider the following padded formulations of our general multiplex networks (adapted here from \cite{ModFAQ,sussman2018matched}).
Letting ${\bf H} \in \mathcal{M}_m^{c}$ and ${\bf G}\in \mathcal{M}_n^{c}$ with $m\leq n$, we consider the following two schemes for ameliorating the differing graph orders.
\begin{enumerate}[label=\roman*.]
    \item (Naive Padding) For each $i\in[c]$, define the weighted adjacency matrices $\widetilde{A}_i\in \R^{m\times m}$ and $\widetilde{B}_i\in \R^{n\times n}$ via 
 \[
 \widetilde{A}_i(u,v)=
    \begin{dcases}
        1 & \text{if u,v}\in V(H_i), \text{ and \{u,v\}}\in E(H_i); \\
        0 & \text{if u,v}\in V(H_i), \text{ and \{u,v\}}\notin E(H_i); \\
        0 & \text{if u or v}\in [m]\setminus V(H_i); \\
    \end{dcases}
\]   
 \[
 \widetilde{B}_i(u,v)=
    \begin{dcases}
        1 & \text{if u,v}\in V(G_i), \text{ and \{u,v\}}\in E(G_i); \\
        0 & \text{if u,v}\in V(G_i), \text{ and \{u,v\}}\notin E(G_i); \\
        0 & \text{if u or v}\in [n]\setminus V(G_i); \\
    \end{dcases}
\] 
Denote ${\bf \widetilde A}=(\widetilde{A}_1,\widetilde{A}_2,\cdots,\widetilde{A}_c)$ and 
${\bf \widetilde B}=(\widetilde{B}_1,\widetilde{B}_2,\cdots,\widetilde{B}_c)$.

    \item (Centered Padding) For each $i\in[c]$, define the weighted adjacency matrices $\widehat{A}_i\in \R^{m\times m}$ and $\widehat{B}_i\in \R^{n\times n}$ via 
\begin{align}
\label{eq:cenpad}
 \widehat{A}_i(u,v)&=
    \begin{dcases}
        1 & \text{if u,v}\in V(H_i), \text{ and \{u,v\}}\in E(H_i); \\
        -1 & \text{if u,v}\in V(H_i), \text{ and \{u,v\}}\notin E(H_i); \\
        0 & \text{if u or v}\in [m]\setminus V(H_i); \\
    \end{dcases}\\
 \widehat{B}_i(u,v)&=
    \begin{dcases}
        1 & \text{if u,v}\in V(G_i), \text{ and \{u,v\}}\in E(G_i); \\
        -1 & \text{if u,v}\in V(G_i), \text{ and \{u,v\}}\notin E(G_i); \\
        0 & \text{if u or v}\in [n]\setminus V(G_i); \\
    \end{dcases}\notag
\end{align}
Denote ${\bf \widehat A}=(\widehat{A}_1,\widehat{A}_2,\cdots,\widehat{A}_c)$ and 
${\bf \widehat B}=(\widehat{B}_1,\widehat{B}_2,\cdots,\widehat{B}_c)$.
\end{enumerate}
\noindent
The \emph{Naive Multiplex Graph Matching Problem} (nMGMP) is then defined as finding an element $P\in\Pi_{n}$ in 
\begin{equation}
\label{eq:nmgmp}
\argmin_{P\in\Pi_{n}}\sum_{i=1}^{c}\|(\widetilde{A}_i\oplus \textbf{0}_{n-m})P-P\widetilde{B}_i\|^2_{F}
\,=\argmin_{P\in\Pi_{n}}\sum_{i=1}^{c}-\text{tr}((\widetilde{A}_i\oplus \textbf{0}_{n-m})P\widetilde{B}_iP^T),
\end{equation}
where $\textbf{0}_{n-m}$ is the $n-m\times n-m$ matrix of all $0$'s. 
The formulation in Eq.\@ (\ref{eq:nmgmp}) effectively seeks to maximize the number of common edges between the multiplex template and multiplex background, where all edges across all channels are weighted equally (see \cite{ModFAQ,bento2018family} for the monoplex analogue).
The \emph{Centered Multiplex Graph Matching Problem} (cMGMP) is defined as finding an element $P\in\Pi_{n}$ in 
\begin{equation}
\label{eq:cmgmp}
\argmin_{P\in\Pi_{n}}\sum_{i=1}^{c}\|(\widehat{A}_i\oplus \textbf{0}_{n-m})P-P\widehat{B}_i\|^2_{F}
\,=\argmin_{P\in\Pi_{n}}\sum_{i=1}^{c}-\text{tr}((\widehat{A}_i\oplus \textbf{0}_{n-m})P\widehat{B}_iP^T).
\end{equation}
If for each $i$, we have that $V(G_i)=[n]>[m_i]=V(H_i)$, then the formulation in Eq.\@ (\ref{eq:cmgmp}) effectively seeks to minimize the number of disagreements (edge mapped to non-edge and vice versa) induced between the background and the matched subgraphs in the template, where all disagreements across all channels are weighted equally.
Given this interpretation, the appropriate padding schemes to deploy in practice depends on the underlying problem assumptions and setting.
\begin{rem}
\label{app:rem1}
\emph{
Our formulation of the Multiplex GMP is (assuming channels of equal order across ${\bf A}$ and ${\bf B}$)
\begin{equation*}
\argmin_{P\in\Pi_{n}}\sum_{i=1}^{c}\|{A}_iP-P{B}_i\|^2_{F}.
\end{equation*}
rather than a formulation weighting the matching in each channel via
\begin{equation*}
\argmin_{P\in\Pi_{n}}\sum_{i=1}^{c}\lambda_i\|{A}_iP-P{B}_i\|^2_{F},
\end{equation*}
for $\lambda_i>0$.
In our subgraph detection setting, we have found that
 $\lambda_i=1$ works suitably well; moreover, this weights each edge in each template channel equally, which may be desirable.
 In the case that one or more channels is more informative or of higher import than the others, then choosing appropriate $\lambda$'s to overweight the matching in those channels may be desirable.
}
\end{rem}

\section{Multiplex Graph Matching Matched Filters}
\label{sec:mgmmf}
Given $\mathcal{A}$, a multiplex graph matching algorithm designed to approximately solve Eqs. (\ref{eq:nmgmp}--\ref{eq:cmgmp}), our multiplex graph matching matched filter (M-GMMF), generalizing the monoplex filtering setting of \cite{sussman2018matched}, proceeds as in Algorithm \ref{alg:mgmmf}.
Note that in our experiments (and in the pseudocode below), we make use of $\mathcal{A}=\texttt{MFAQ}$ (see Algorithm \ref{alg:mfaq} in Appendix \ref{app:mfaq}), but we stress that our approach can utilize any suitable $\mathcal{A}$ equally well.
\begin{algorithm}[h!]
  \begin{algorithmic}
    \STATE \textbf{Input}: Multiplex graphs ${\bf H}\in\mathcal{M}^c_m$ and ${\bf G}\in\mathcal{M}^c_n$ with $m<n$; padding regime; tolerance $\epsilon\in\mathbb{R}>0$; restarts $N$
\vspace{3mm}
\STATE {\bf 1.} Pad ${\bf H}$ and ${\bf G}$ accordingly; in the naive (resp., centered) padding regime, the padded ${\bf H}$ is denoted via $\tba$ (resp., $\hba$), and the padded ${\bf G}$ via $\tbb$ (resp., $\hbb$);
\FOR {$k=1,2,\cdots,N$,}
\STATE {\bf 2.} $P^{(0)}\leftarrow \alpha J_n/n+(1-\alpha) P$ where $P\sim\text{Unif}(\Pi_n)$ and $\alpha\sim$Unif[0,1];
\STATE {\bf 3.} In the naive (resp., centered) padding regime, $P^*_k\leftarrow\texttt{MFAQ}(\tba,\tbb,P^{(0)},\epsilon)$) (resp., $P^*_k\leftarrow\texttt{MFAQ}(\hba,\hbb,P^{(0)},\epsilon)$);
\ENDFOR
\STATE{\bf 4.} Rank the matchings $\{P^*_1,P^*_2,\cdots,P^*_N\}$ by increasing value of the multiplex graph matching objective function, Eq. \ref{eq:nmgmp} or \ref{eq:cmgmp}, depending on the padding regime selected;
\STATE \textbf{Output}: Ranked list $(P^*_{(1)},P^*_{(2)},\cdots,P^*_{(N)})$ of matchings, aligning multiplex template ${\bf H}$ to background ${\bf G}$.
\end{algorithmic}
\caption{M-GMMF}
\label{alg:mgmmf}
\end{algorithm}

Effectively, the M-GMMF algorithm uses the multiplex template (and algorithm $\mathcal{A}$) to search $\Pi_n$ for suitable solutions aligning $\bh$ to $\bg$.
The multiple restarts in Step 2. of the procedure are needed in the case of $\mathcal{A}=\texttt{M-FAQ}$, as in that setting the objective function is relaxed to an indefinite quadratic program with myriad local minima in the feasible region; these restarts aim to precisely counteract the presence of these local minima by broadly searching the feasible region for a global minimum.
For approximate combinatorial $\mathcal{A}$, the restarts may be appropriate as well, while for continuous, convex relaxation algorithms (see, for example, \cite{bento2018family}), this step may not be necessary.

Note that code implementing the above \texttt{M-GMMF} and \texttt{M-FAQ} procedures can be downloaded as part of our \texttt{R} package, \texttt{iGraphMatch}, which is available on \texttt{CRAN} or can be downloaded at \url{https://github.com/dpmcsuss/iGraphMatch}.

\subsection{Multiplex Matchability}
\label{ssec:MMbility}

In \cite{sussman2018matched}, the authors considered an error model wherein the template $\bh$ is an errorful induced subgraph of the background $\bg$ in the monoplex setting.
The aim of the Monoplex-GMMF approach then was to recover the vertices in $\bg$ corresponding to $\bh$.
Can we recover the analogous results in the multiplex setting?
To frame and attack this problem statistically, we consider the following error model which we will use to generate a multiplex background graph ${\bf G}\in\mn$ and a multiplex template ${\bf H}\in\mm$ with $m<n$.
\bde[See \cite{arroyo2018maximum}]
Consider a graph $G$ with $V(G)\subset [n]$.
Let the centered, padded adjacency matrix (as in Eq. (\ref{eq:cenpad})) of $G$ be denoted 
$\widehat A\in\mathbb{R}^{n\times n}$.
Let $E\in[0,1]^{n\times n}$ be a symmetric, hollow matrix.  
The graph-valued random variable $\mathcal{E}(G)$ with vertex set equal to $V(G)$ and random centered, padded adjacency matrix $\widehat A_{G,E}$, which models passing $G$ through an errorful channel $E$, is defined as follows.
For each $\{i,j\}\in\binom{[n]}{2}$, 
$$\widehat A_{G,E}(i,j)=\widehat A(i,j)\cdot (1-2X(i,j)),$$
where $X(i,j)\stackrel{ind.}{\sim}Bern(E(i,j))$.
\ede
\noindent The two generative models we then consider are defined via:
\begin{enumerate}
\item[i.] (\emph{Single Channel Source, Error Multiplex, abbreviated ME}) 
There is a single non-random background source graph $W\in\gn$ and non-random source template $T=W[m]\in\gm$, and two multi-channel errorful filters ${\bf E}^{(1)}=(E^{(1)}_1,\ldots,E^{(1)}_c)$, with each $E^{(1)}_i$ acting on $W$, and ${\bf E}^{(2)}=(E^{(2)}_1,\ldots,E^{(2)}_c)$, with each $E^{(2)}_i$ acting on $T$.
We observe 
${\bf G}=(\mathcal{E}^{(1)}_1(W),\ldots,\mathcal{E}^{(1)}_c(W))$ as the multiplex background and 
${\bf H}=(\mathcal{E}^{(2)}_1(T),\ldots,\mathcal{E}^{(2)}_c(T))$ as the multiplex template.
By assumption, the errorful filters act independently across channels within ${\bf G}$ and ${\bf H}$, and independently across ${\bf G}$ and ${\bf H}$.
In this model, by construction each $|V(H_i)|=[m]$ and each $|V(G_i)|=[n]$.

\item[ii.] (\emph{Single Channel Errors, Source Multiplex, abbreviated MS}) The non-random background and non-random template source graphs are multiplex.  
To wit, let ${\bf T}\in\mm$ and ${\bf W}\in\mn$ satisfy the following:  For each $i\in[c]$, let 
$\widehat{\bf C}$ and 
$\widehat{\bf D}$ be the centered paddings of 
${\bf T}$ and ${\bf W}$ respectively.
We assume then that
$\hc_i=\hd_i[m]$ (i.e., $\hc_i$---the padded adjacency matrix of $T_i$---is the $m\times m$ principal submatrix of $\hd_i$---the padded adjacency matrix of $W_i$).
There are two multi-channel errorful filters: ${\bf E}^{(1)}=(E^{(1)}_1,\ldots,E^{(1)}_c)$ and ${\bf E}^{(2)}=(E^{(2)}_1,\ldots,E^{(2)}_c)$.
For each $i\in[c]$,  $E^{(1)}_i\in\mathbb{R}^{n\times n}$ acts on $W_i$, and $E^{(2)}_i\in\mathbb{R}^{m\times m}$ acts on $T_i$.
We observe 
${\bf G}=(\mathcal{E}^{(1)}_1(W_1),\mathcal{E}^{(1)}_2(W_2)\ldots,\mathcal{E}^{(1)}_c(W_c))$ as the multiplex background and 
${\bf H}=(\mathcal{E}^{(2)}_1(T_1),\mathcal{E}^{(2)}_2(T_2),\ldots,\mathcal{E}^{(2)}_c(T_c))$ as the multiplex template.
As above, the errorful filters act independently across channels within ${\bf G}$ and ${\bf H}$, and independently across ${\bf G}$ and~${\bf H}$.
Note that if the template (resp., background) channels have non-identical vertex sets, then this will be preserved in the errorful template (resp., background).
\end{enumerate}
It may be convenient to view $T$ and $W$ (resp., ${\bf T}$ and ${\bf W}$) as realizations from graph-valued random variables in the ME (resp., MS) model. 
In this case, we will assume the actions of the errorful filters on $T$ and $W$ (resp., ${\bf T}$ and ${\bf W}$) are also independent of the random $T$ and $W$ (resp., ${\bf T}$ and ${\bf W}$).

Considering the models above, in order for our M-GMMF approach to possibly recover the \emph{true} errorful induced subgraph of $\bg$ corresponding to $\bh$, we need for the global minimum of the Multiplex GMP to be in 
$\mathcal{P}_{m,n}:=\{I_m\oplus P: P\in\Pi_{n-m}\}$.
This is the multiplex analogue of graph matchability, i.e., uncovering conditions under which oracle graph matching will recover a latent vertex alignment.
Here, that alignment is represented by $\bh$ being an errorful version of $\bg[m]$; see, for example, \cite{ma1,ma2,ma3,ma4,ma5,ma6,ma7,ma8,ma9,ma10} for a litany of graph matchability results in the monoplex setting.

\subsection{MS model matchability}
\label{sec:MSmatch}

In this section, we will explore the benefit of considering multiplex versus monoplex networks when considering template matchability in the MS model.
We note here that while the formal theory underlying the matchability results in the multiplex setting differs only slightly from the monoplex setting of \cite{sussman2018matched}, we stress that the end results demonstrate the utility of considering multiple channels.

In the MS model, let ${\bf T}\in\mathcal{M}_m^c$ and ${\bf W}\in\mathcal{M}_n^c$ be the respective template and background source graphs, with respective centered, padded adjacency matrices given respectively by $\widehat{\bf C}$ and $\widehat{\bf D}$ satisfying $\widehat{C}_i=\widehat{D}_i[m]$ for all $i\in[c]$.
Assume that the errorful filters satisfy for each $i\in[c]$, $E_i^{(1)}=q_i J_n$ and $E_i^{(2)}=s_i J_m$ (where $s_i=s_i(n)$ and $q_i=q_i(n)$ are allowed to vary with $n$).
If $c=1$, and $s_1=q_1=1/2$, then the observed background and template are effectively independent ER$(n,1/2)$ and ER$(m,1/2)$ networks, respectively.
It is immediate then that the optimal permutation aligning the background to the template will almost surely not be in $\mathcal{P}_{m,n}$. 

Consider now $c>1$.
Let $\mathcal{B}=\{i\in[c]\text{ s.t. }s_i\text{ or }q_i=1/2\}$; these ``bad'' channels act to obfuscate the latent alignment between $\widehat{\bf C}$ and $\widehat{\bf D}$ by effectively whitening the signal present in the alignment within the channels.  
Suppose that there exist constants $\alpha\leq1$, $\beta>0$, and $n_0\in\mathbb{Z}>0$ such that for all $n>n_0$, $m=m(n)$ satisfies $m^{\alpha}>\beta\log n$.
For each $m$, denote the set of permutations that permute exactly $k$ labels of $[m]$ by $\Pi_{n,m,k}$, and 
for each $P\in \Pi_n$ (with associated permutation $\sigma_p$), define 
\begin{align}
\label{eq:delp}
\Delta_P=\left\{ \{i,j\}\in\binom{[m]}{2}\text{ s.t. }\{i,j\}\neq\{\sigma_p(i),\sigma_p(j)\}\right\},
\end{align} and for each $i\in[c]$, define
\begin{align*}
\Delta_P^{(i,1)}&=\{\,\{j,\ell\}\in\Delta_P\text{ s.t. }0\neq\hc_i(j,\ell)\neq\hd_i(\sigma_p(j),\sigma_p(\ell))\neq 0  \};\\
\Delta_P^{(i,2)}&=\{\,\{j,\ell\}\in\Delta_P\text{ s.t. }0\neq\hc_i(j,\ell)\neq\hd_i(\sigma_p(j),\sigma_p(\ell))=0  \}.
\end{align*}
Suppose that there exists an $n_1>0$ such that for all $n>n_1$, we have that for all $k\in[m=m(n)]$ and all $P\in\Pi_{n,m,k}$,
\begin{align}
\label{eq:msneed}
\sum_{i\in[c]\,\setminus\,\mathcal{B}} \left(2|\Delta_P^{(i,1)}|+|\Delta_P^{(i,2)}|\right)(1-2s_i)(1-2q_i)\geq k\sqrt{\frac{672 m^{1+\alpha}c}{\beta}}.
\end{align}
Letting $\widehat{\bf A}$ and $\widehat{\bf B}$ be the padded, centered adjacency matrices of ${\bf H}$ and ${\bf G}$ respectively (the errorful ${\bf T}$ and ${\bf W}$), for $n>\mathfrak{n}=\max(n_0,n_1)$ we have that
\begin{align}
\label{eq:bndprob}
\p\left(\argmin_{P\in\Pi_{n}}\sum_{i=1}^{c}\|(\widehat{A}_i\oplus \textbf{0}_{n-m})P-P\widehat{B}_i\|^2_{F}\not\subset \mathcal{P}_{m,n} \right)\leq2n^{-2},
\end{align}
(see Appendix \ref{app:MS} for proof of this bound).

Exploring Eq. (\ref{eq:msneed}) in the ER setting further, we consider the following setup.
If $c=c(n)\leq n$, and for each $i\in[c]$, $W_i\sim ER(n,p_i)$ with $p_i=p_i(n)\leq1/2$, 
then for each $P\in\Pi_{n,m,k}$, a simple application of McDiarmid's inequality (see Appendix \ref{app:rem1}) yields that 
$$|\Delta_P^{(i,1)}|\in(\,|\Delta_P|p_i(1-p_i),3|\Delta_P|p_i(1-p_i)\,),$$ 
with probability at least 
\begin{align}
\label{eq:mcd3}
1-2\text{exp}\left\{\frac{-2 |\Delta_P|p_i^2(1-p_i)^2 }{8}\right\}.
\end{align}
Note that if $m>6$, then $mk/3\leq|\Delta_P|\leq mk$, so that with probability at least Eq. (\ref{eq:mcd3}),
$$
|\Delta_P^{(i,1)}|\in(1/3,3)\cdot mkp_i(1-p_i).
$$
Suppose that $\alpha<1$, and that there exists an $n_2>0$ such that for all $n>n_2$, we have that for all $i\in[c]$, $mp_i^2\geq 384\log n$, and 
\begin{align}
\label{eq:ERneedMS}
\sum_{i\in[c]\,\setminus\,\mathcal{B}} p_i(1-2s_i)(1-2q_i)\geq\sqrt{\frac{6048m^{\alpha-1}c}{\beta}}.
\end{align}
For $n>\max(n_2,n_0)$, we then have
\begin{align}
\label{eq:erms}
\p&\bigg(\underbrace{\argmin_{P\in\Pi_{n}}\sum_{i=1}^{c}\|(\widehat{A}_i\oplus \textbf{0}_{n-m})P-P\widehat{B}_i\|^2_{F}\not\subset \mathcal{P}_{m,n}}_{:=\mathcal{A}_n} \bigg)\leq 4n^{-2}.
\end{align}
For proof of Eq.\@ (\ref{eq:erms}), see Appendix \ref{app:erms}.

We have thus proven the following theorem:
\begin{theorem}
\label{thm:goodvbad}
With setup as above, 
suppose that $\alpha<1$, and $p_i=p$ is a fixed constant that does not vary with $n$.
Further suppose that $s_i=s<1/2$, and for $c_1$ channels $q_i=q<1/2$, and for $c_2=c-c_1-|\mathcal{B}|$ channels $q_i=1-q>1/2$ (where $c_1=c_1(n)$, $c_2=c_2(n)$, $s=s(n)$ and $q=q(n)$ are allowed to vary with $n$).
Then there exist constants $\gamma,\xi>0$, and $n_2\in\mathbb{Z}>0$ such that 
if for all $n>n_2$,
\begin{align}
\label{eq:goodvbad}
mp_i^2\geq \xi \log n\text{ for all }i\in[c]\notag,\\
(1/2-s)(1/2-q)(c_1-c_2)>\gamma\sqrt{m^{\alpha-1}c},
\end{align}
then for $n>\max(n_0,n_2)$, $\p(\mathcal{A}_n)\leq 4n^{-2}.$
If $s$, $q$, $c$, $c_1$ and $c_2$ are fixed constants that do not vary with $n$, we need only require $c_1>c_2$ rather than Condition (\ref{eq:goodvbad}).
\end{theorem}
\subsubsection{Strength in numbers}
\label{sec:strnum}

Consider $c_2=|\mathcal{B}|=0$ in Theorem \ref{thm:goodvbad}.
Condition (\ref{eq:goodvbad}) then reduces to
\begin{align}
\label{eq:goodvbad1}
(1/2-s)(1/2-q)\sqrt{c}>\gamma\sqrt{m^{\alpha-1}},
\end{align}
and large values (i.e., close to $1/2$) of $s$ and $q$ can be mitigated by choosing an appropriately large $c$; effectively, multiple channels can amplify the weak signal present in each individual channel.

We explore this further in the following experiment.  
We will look at two different cases specifically, when $m=n$ and when $m<n$. First, considering $n=m=100$ (to mitigate possible effects of template order on matching accuracy), 
we let ${\bf G},{\bf H}\in \mathcal{M}^{c}_{100}$ for $c$ ranging over $\{1,2,\cdots,10\}$.
For each $i\in[c]$, we have that 
$(G_i,H_i)\sim\mathrm{ER}(100,0.5,\rho)$ (so that $G_i$ and $H_i$ are marginally ER(100,0.5) and edges across graphs are independent except that for each $\{j,k\}\in\binom{[100]}{2}$, we have that corr$(\mathds{1}\{\{j,k\}\in E(G_i)\},\mathds{1}\{\{j,k\}\in E(H_i)\})=\rho$).
\begin{figure}[t!]
    \centering
    \includegraphics[width=0.7\textwidth]{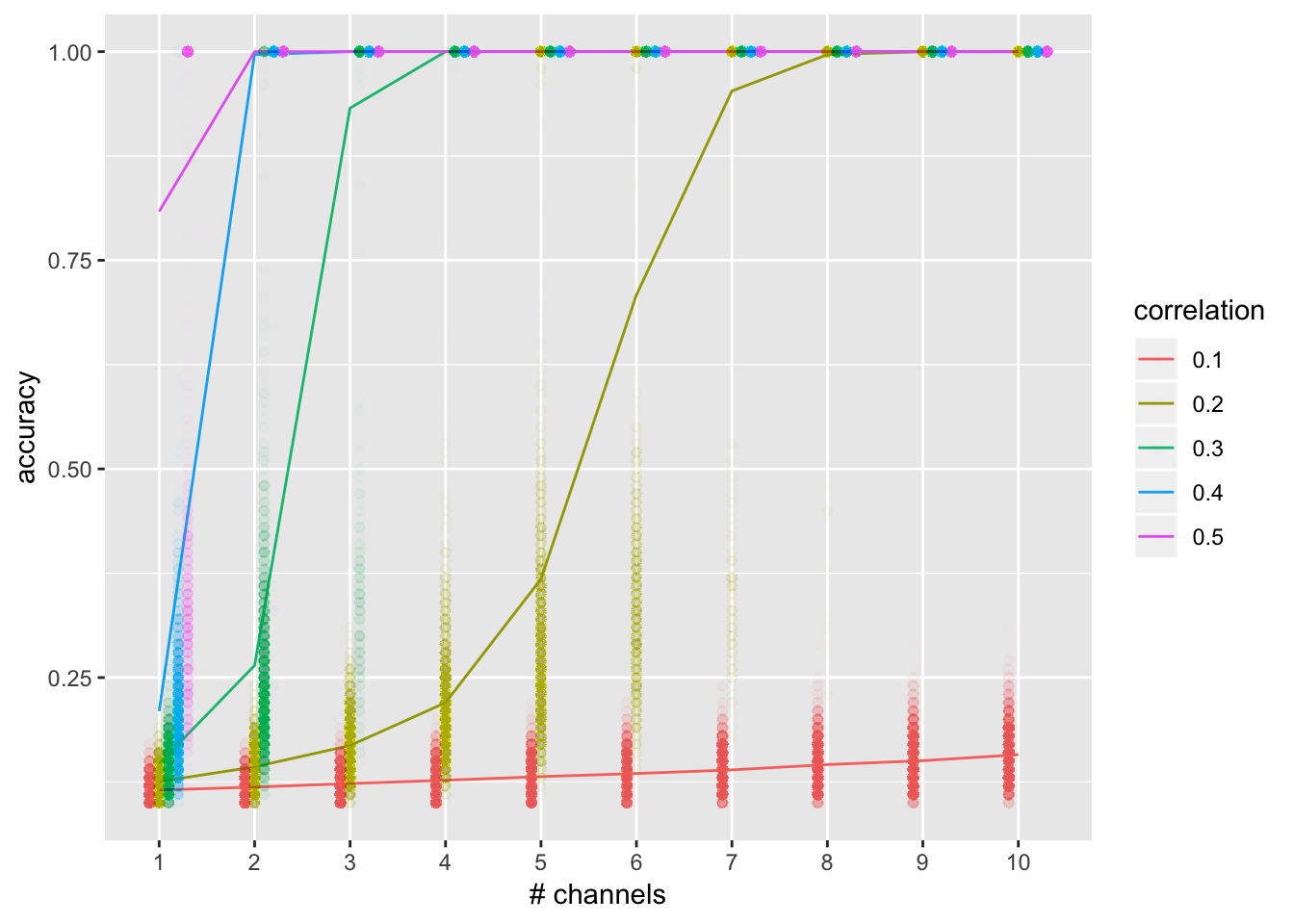} 
    \caption{
  Considering $n=m=100$, 
we let ${\bf G},{\bf H}\in \mathcal{M}^{c}_{100}$ (for $c$ ranging over $\{1,2,\cdots,10\}$). 
For $i\in[c]$, we have that 
$(G_i,H_i)\sim$ ER$(100,0.5,\rho)$.
Utilizing $s=10$ seeded vertices, we match ${\bf G}$ and ${\bf H}$ using \texttt{MFAQ} (Algorithm \ref{alg:mfaq}). 
In red (resp., olive, green, blue, purple) we plot the results for $\rho=0.1$ (resp., $\rho=0.2$, $\rho=0.3$, $\rho=0.4$, $\rho=0.5$).
The partially transparent points visualize the accuracy distribution and correspond to individual Monte Carlo replicates.}
    \label{fig:goodvbad}
\end{figure}
Within this model, the channels are endowed with a natural vertex correspondence across $G_i$ and $H_i$, namely the identity mapping.
Note that in the $W_i\sim$ER$(n,p_i)$ MS model setting, we have that 
$\text{Cov}(\mathds{1}\{\{j,k\}\in E(G_i)\},\mathds{1}\{\{j,k\}\in E(H_i)\})=p_i(1-p_i)(1-2s_i)(1-2q_i),$
so that the correlation between edges in $G_i$ and $H_i$ can be made positive or negative with judiciously chosen $s_i$ and $q_i$. Considering $\rho$ varying over 
$\{0.1,0.2,0.3,0.4,0.5\}$, we match ${\bf G}$ and ${\bf H}$ using \texttt{M-FAQ} (Algorithm \ref{alg:mfaq} using $s=10$ seeded vertices \cite{ModFAQ}). Results are plotted in Figure \ref{fig:goodvbad}. In Figure \ref{fig:goodvbad}, we plot the mean matching accuracy (i.e., the fraction of vertices whose latent alignment is recovered correctly) of \texttt{M-FAQ} versus $c$, averaged over 2000 Monte Carlo replicates.
For each choice of parameters, we also plot (via the partially transparent points) the accuracy distribution corresponding to the MC replicates. 
In red (resp., olive, green, blue, purple) we plot the results for $\rho=0.1$ (resp., $\rho=0.2$, $\rho=0.3$, $\rho=0.4$, $\rho=0.5$).
From Figure \ref{fig:goodvbad}, we see the expected relationship:  in low correlation settings where \texttt{M-FAQ} is unable to align the monoplex graphs, this can often be overcome by considering $c>1$.
Indeed, in all cases, save $\rho=0.1$, perfect matching is achieved using $c\geq 8$ channels.

\begin{figure}[t!]
\begin{subfigure}{.5\textwidth}
  \centering
  \includegraphics[width=.9\linewidth]{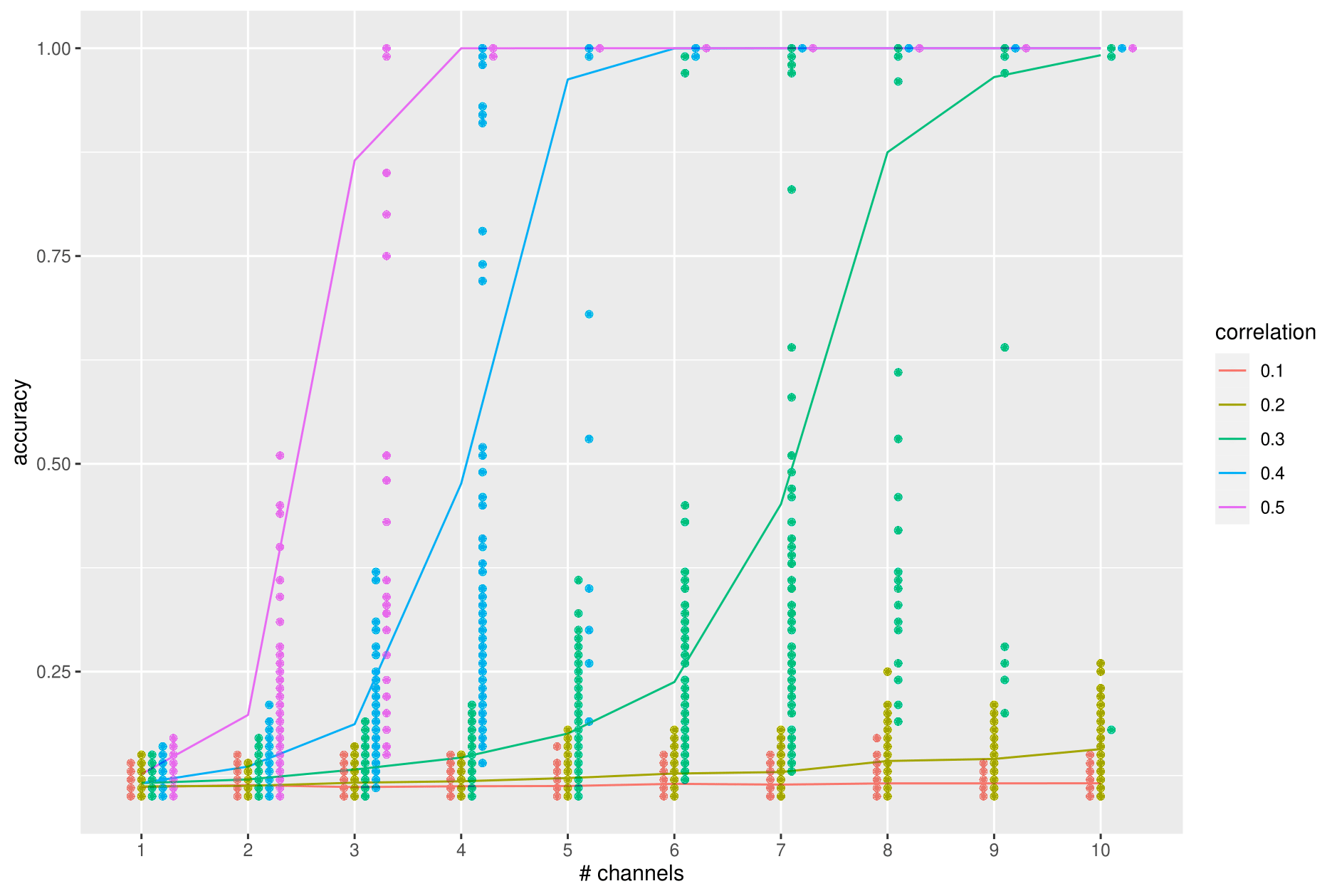}
  \caption{Naive padding; $(\widetilde{G}_i,\widetilde{H}_i)\sim\mathrm{ER}(500,0.5,\rho)$,\\  where $\rho\in
\{0.1,0.2,0.3,0.4,0.5\}$, $i\in[10]$.}
\end{subfigure}%
\begin{subfigure}{.5\textwidth}
  \centering
  \includegraphics[width=.9\linewidth]{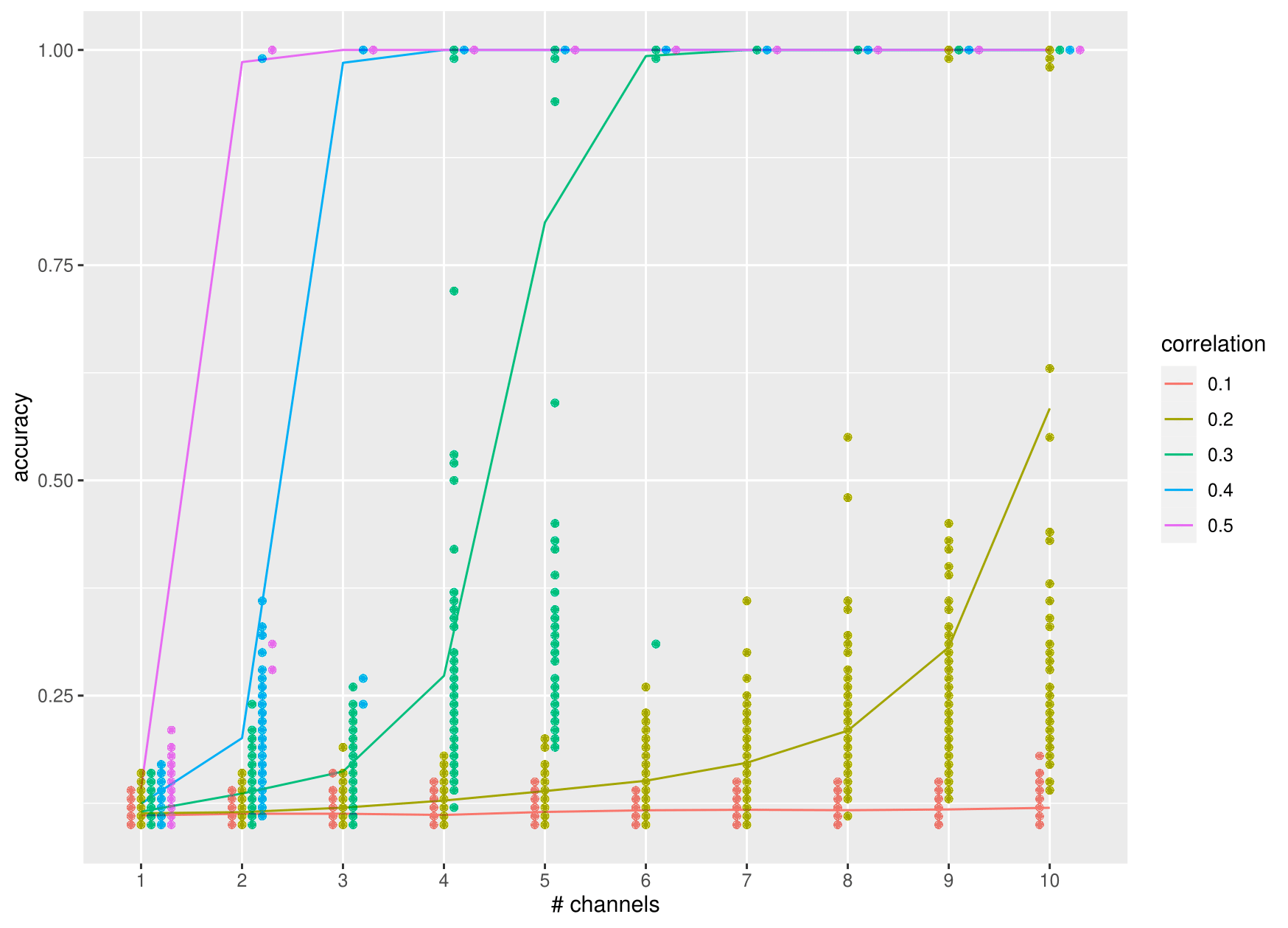}
  \caption{Centered padding; $(\widehat{G}_i,\widehat{H}_i)\sim\mathrm{ER}(500,0.5,\rho)$,\\  where $\rho\in
\{0.1,0.2,0.3,0.4,0.5\}$, $i\in[10]$.}
\end{subfigure}
\caption{Under the same MS model setting as in Figure \ref{fig:goodvbad}, we fix $n=500$, $m=100$ and we apply naive padding (left panel) and centered padding (right panel) in $({\bf G},{\bf H})\in(\mathcal{M}^{10}_{500},\mathcal{M}^{10}_{100})$.
 We plot the matching accuracy (averaged over 100 Monte Carlo replicates) obtained by \texttt{M-FAQ} (with $s=10$ seeded vertices) versus the number of channels $c$.
In red (resp., olive, green, blue, purple) we plot the results for $\rho=0.1$ (resp., $\rho=0.2$, $\rho=0.3$, $\rho=0.4$, $\rho=0.5$).
The partially transparent points visualize the accuracy distribution and correspond to individual Monte Carlo replicates.}
\label{fig:goodvbad_rhovary}
\end{figure}

\begin{figure}[t!]
\begin{subfigure}{.5\textwidth}
  \centering
  \includegraphics[width=.9\linewidth]{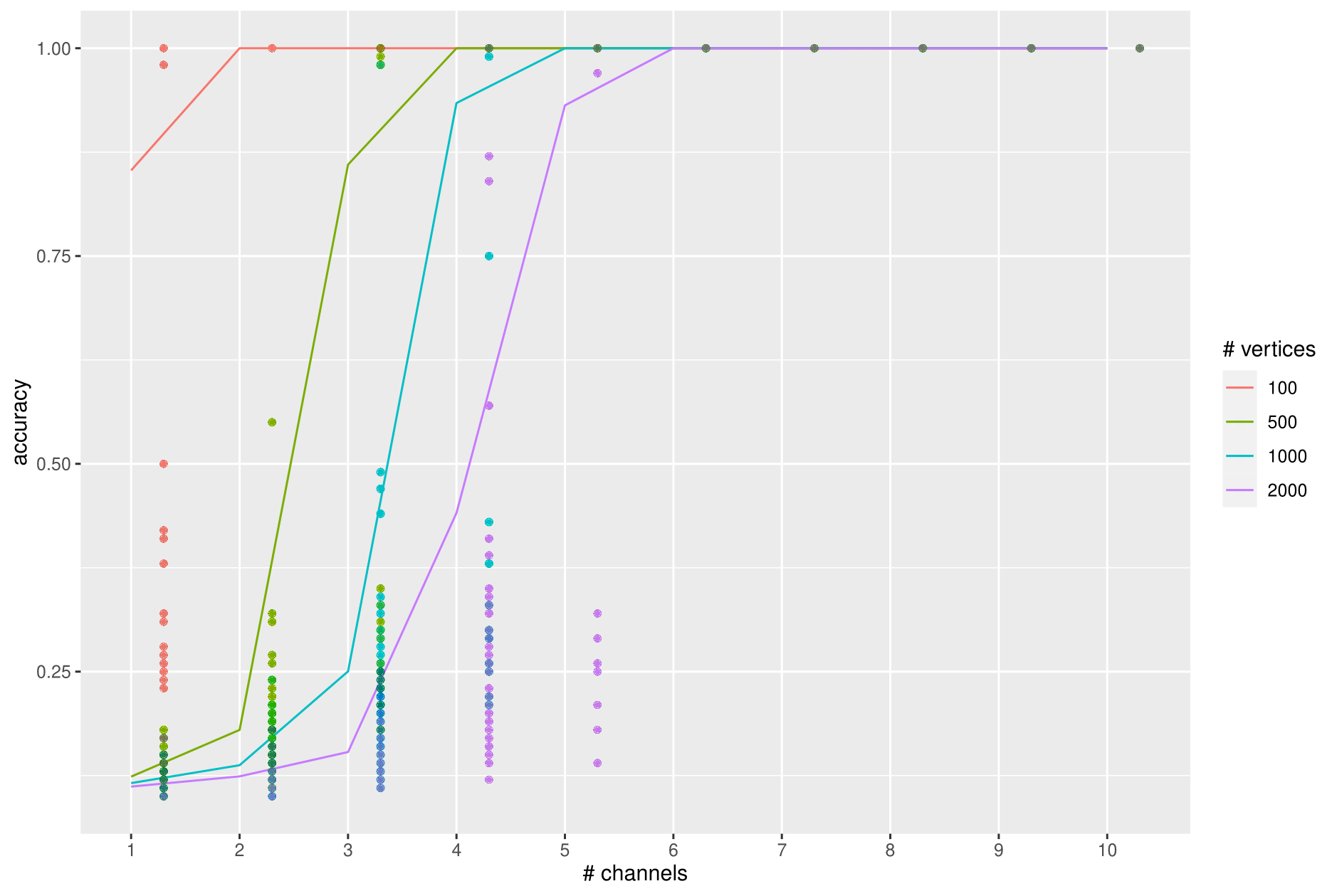}
  \caption{Naive padding; $(\widetilde{G}_i,\widetilde{H}_i)\sim\mathrm{ER}(n,0.5,0.5)$,\\  where $n\in
\{100,500,1000,2000\}$, $i\in[10]$.}
\end{subfigure}%
\begin{subfigure}{.5\textwidth}
  \centering
  \includegraphics[width=.8\linewidth]{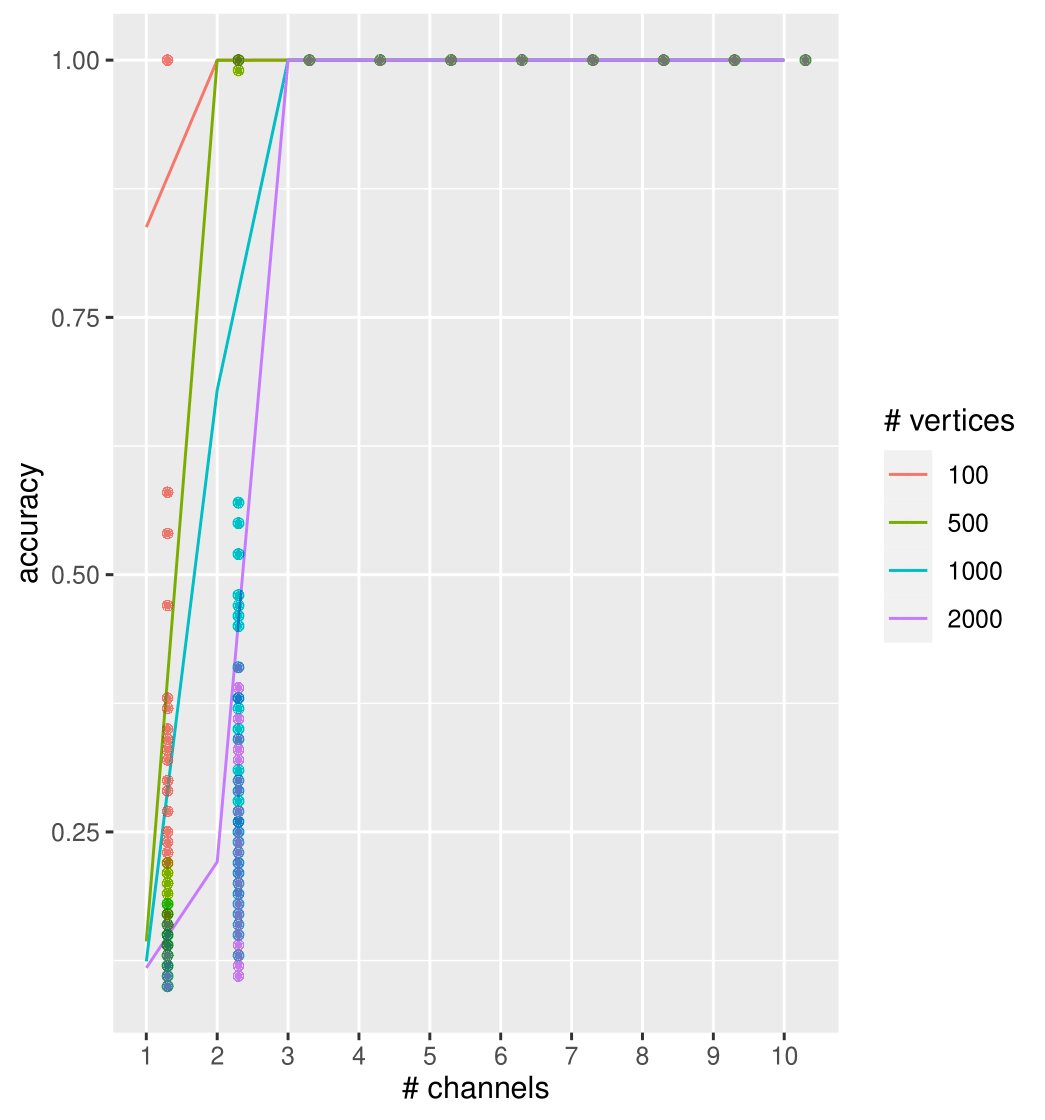}
  \caption{Centered Padding; $(\widehat{G}_i,\widehat{H}_i)\sim\mathrm{ER}(n,0.5,0.5)$,\\  where $n\in
\{100,500,1000,2000\}$, $i\in[10]$.}
\end{subfigure}
\caption{Under the same MS model setting as in Figure \ref{fig:goodvbad}, we fix $m=100$ and we consider $n$ ranging over $\{100,500,1000,2000\}$.
We fix $\rho = 0.5$ and we apply naive padding (left panel) and centered padding (right panel) in $({\bf G},{\bf H})\in(\mathcal{M}^{10}_{n},\mathcal{M}^{10}_{100})$.
We match both pairs ${\bf \widetilde{G}}$, ${\bf \widetilde{H}}$ and ${\bf \widehat{G}}$, ${\bf \widehat{H}}$ using \texttt{MFAQ} (with $s=10$ seeds). 
In red (resp., green, blue, purple) we plot the results for $n=100$ (resp., $n=500$, $n=1000$, $n=2000$). The partially transparent points visualize the accuracy distribution and correspond to individual Monte Carlo replicates.}
\label{fig:goodvbad_nnvary}
\end{figure}

Next, we look at the case when $m<n$. In addition to examining the effect of multiple channels when weak signal is present across channels, we wish to compare the effect of different padding schemes (Naive vs Centered) in terms of the matching accuracy. We analyze the padding scheme's effectiveness first, by varying the values of the correlation $\rho\in\{0.1,0.2,0.3,0.4,0.5\}$ while keeping $n,m$ constant (see Figure \ref{fig:goodvbad_rhovary}) and second, by varying the background size $n\in\{100,500,1000,2000\}$ while the template size $m$ and the correlation $\rho$ remain constant (see Figure \ref{fig:goodvbad_nnvary}). Using the Naive (resp. Centered) padding scheme, we let $({\bf \widetilde G},{\bf \widetilde H})\in \mathcal{M}^{c}_{n}$ (resp. $({\bf \widehat G},{\bf \widehat H})\in \mathcal{M}^{c}_{n}$) for $c$ ranging over $\{1,2,\cdots,10\}$.
Utilizing $s=10$ seeds, we match ${\bf \widetilde G}$ and ${\bf \widetilde H}$ (resp. ${\bf \widehat G}$ and ${\bf \widehat H}$) using \texttt{M-FAQ} (Algorithm \ref{alg:mfaq}). Results are plotted in Figures \ref{fig:goodvbad_rhovary} and \ref{fig:goodvbad_nnvary}. As in Figure \ref{fig:goodvbad}, we plot the mean matching accuracy (i.e., the fraction of vertices whose latent alignment is recovered correctly) of \texttt{M-FAQ} versus $c$, averaged over 100 MC replicates.
For each choice of parameters, we also plot (via the partially transparent points) the accuracy distribution corresponding to the MC replicates. In Figure \ref{fig:goodvbad_rhovary}, in red (resp., olive, green, blue, purple) we plot the results for $\rho=0.1$ (resp., $\rho=0.2$, $\rho=0.3$, $\rho=0.4$, $\rho=0.5$). In Figure \ref{fig:goodvbad_nnvary}, in red (resp., green, blue, purple) we plot the results for $n=100$ (resp., $n=500$, $n=1000$, $n=2000$).

    All figures demonstrate that even though the \texttt{M-FAQ} algorithm is unable to align the monoplex graphs when $c=1$, this can often be overcome by considering $c>1$.
Moreover, Figures \ref{fig:goodvbad_rhovary} and \ref{fig:goodvbad_nnvary}, show that the Centered Padding scheme achieves better matching accuracy between channels than the Naive Padding scheme.
Akin to Figure \ref{fig:goodvbad}, Figure \ref{fig:goodvbad_rhovary} illustrates that the matching accuracy increases as the correlation increases. Finally, in Figure \ref{fig:goodvbad_nnvary}, we observe that the matching accuracy decreases as the ratio between the template size and the background size decreases.

\subsubsection{The good outweighs the bad}
\label{sec:goodvbad}

In this section, we explore the ability of the signal in ``good'' channels to overcome the obfuscating effect of ``bad'' channels.
To wit, consider Condition (\ref{eq:goodvbad}) with $c_2>0$.
We see that if there are enough channels (i.e., $c_1$ is sufficiently large) with positive correlation ($s_i,q_i<1/2$), then the template and background remain matchable even in the presence of (potentially) multiple anti-correlated channels.

We explore this further in the following experiment. As in the previous subsection \ref{sec:strnum}, we study this ``obfuscating" effect for both $m=n$ and $m<n$ cases.
Again consider $n=m=100$, and let ${\bf G},{\bf H}\in \mathcal{M}^{10}_{100}$ (i.e., $c=10$), where for $i\in[10]$ we have that 
$(G_i,H_i)\sim\mathrm{ER}(100,0.5,\rho)$. Under the same setting, we let $n=500$ and we apply Naive Padding in $({\bf G},{\bf H})\in(\mathcal{M}^{10}_{500},\mathcal{M}^{10}_{100})$, so that $(\widetilde{G}_i,\widetilde{H}_i)\sim\mathrm{ER}(500,0.5,\rho)$ for all $i\in[10]$.
\begin{figure}[t!]
\begin{subfigure}{.5\textwidth}
  \centering
  \includegraphics[width=.85\linewidth]{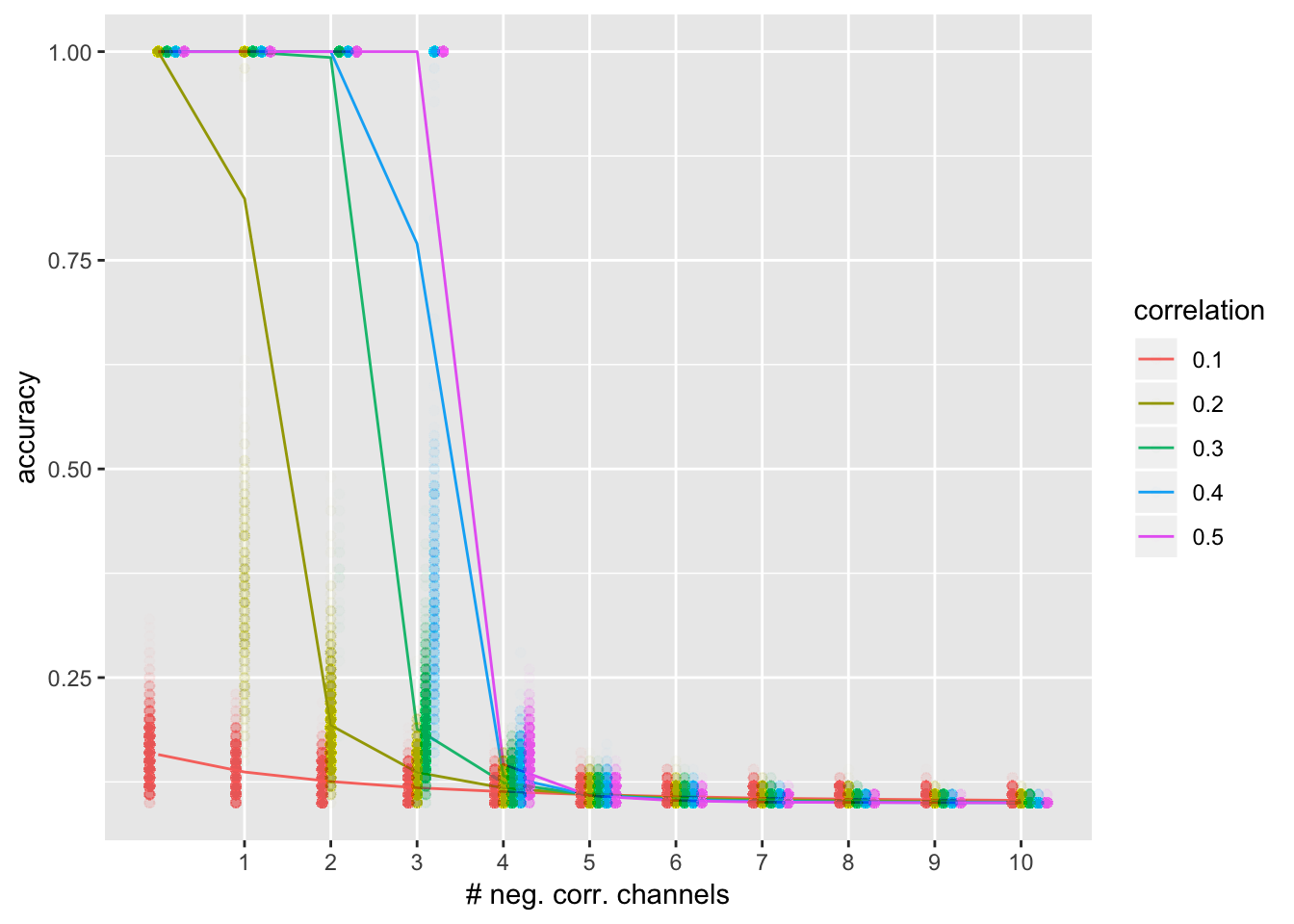}
  \caption{\hspace{1cm} $(G_i,H_i)\sim\mathrm{ER}(100,0.5,\rho)$,\\ where $r\in
\{0.1,0.2,0.3,0.4,0.5\}$, $i\in[10]$.}
\end{subfigure}%
\begin{subfigure}{.5\textwidth}
  \centering
  \includegraphics[width=.85\linewidth]{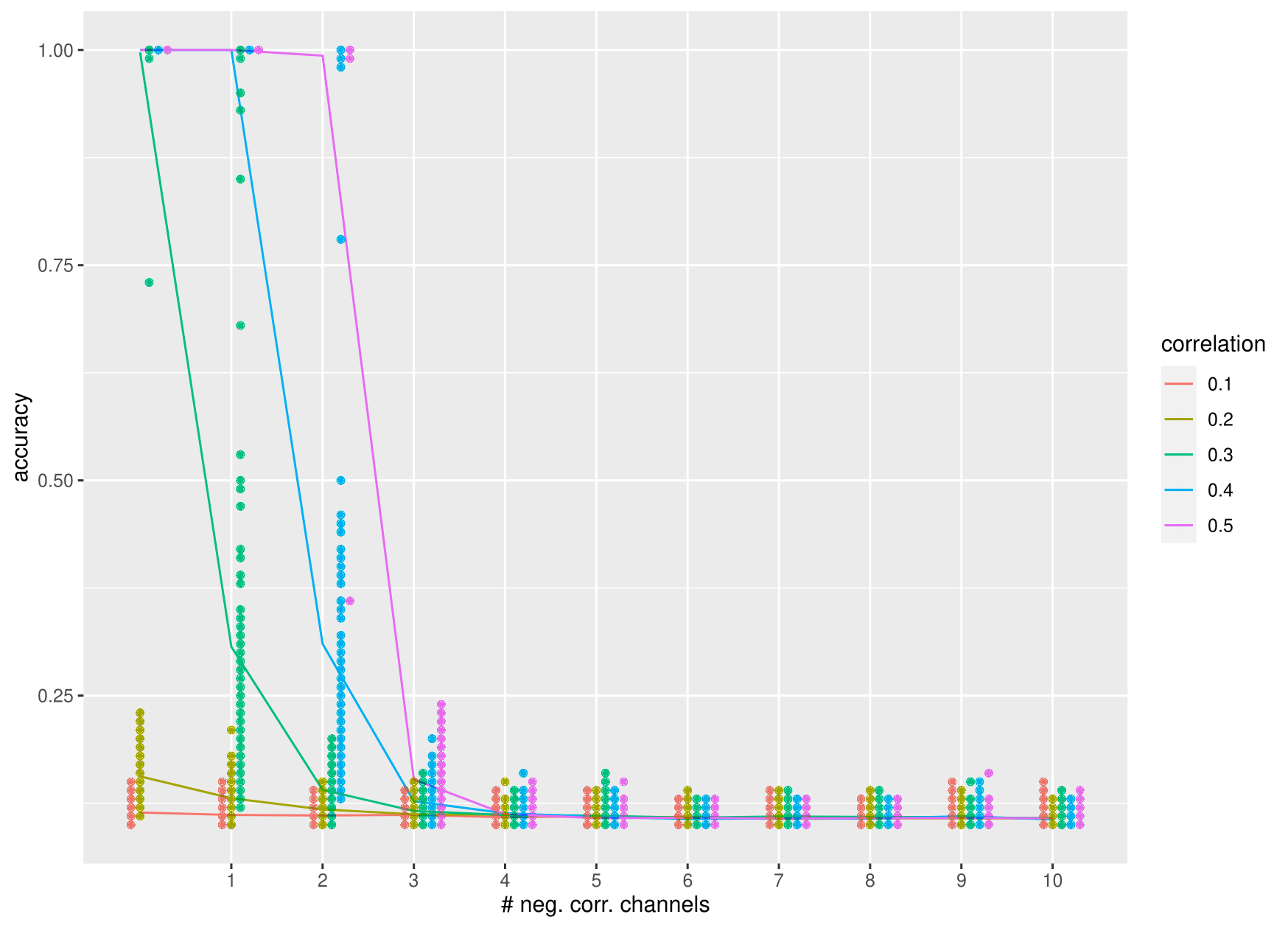}
  \caption{Naive padding; $(\widetilde{G}_i,\widetilde{H}_i)\sim\mathrm{ER}(500,0.5,\rho)$,\\  where $r\in
\{0.1,0.2,0.3,0.4,0.5\}$, $i\in[10]$.}
\end{subfigure}
\caption{We consider $m=100$, and we let ${\bf G},{\bf H}\in \mathcal{M}^{10}_{100}$ (i.e., $c=10$) (left panel). We also implement the Naive Padding scheme, and we let ${\bf \widetilde G},{\bf \widetilde H}\in \mathcal{M}^{10}_{500}$ (right panel).
Considering $\rho$ to take two possible values: $\rho=r$ for $c_g$ channels, or
$\rho=-r$ for $c_b=c-c_g$ channels, where $r$ varies in 
$\{0.1,0.2,0.3,0.4,0.5\}$.
We plot the matching accuracy (averaged over 2000 (left panel) and 100 (right panel) Monte Carlo replicates) obtained by \texttt{M-FAQ} (with 10 seeds) versus $c_b$.
In red (resp., olive, green, blue, purple) we plot the results for $r=0.1$ (resp., $r=0.2$, $r=0.3$, $r=0.4$, $r=0.5$).
The partially transparent points visualize the accuracy distribution and correspond to individual Monte Carlo replicates.
}
\label{fig:goodvbad2}
\end{figure}

\begin{figure}[t!]
\begin{subfigure}{.5\textwidth}
  \centering
  \includegraphics[width=.7\linewidth]{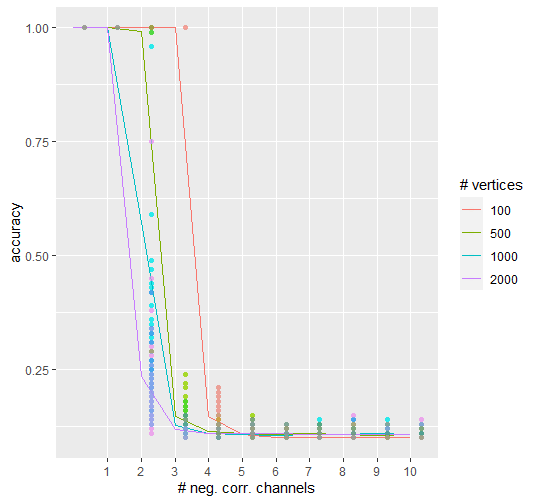}
  \caption{Naive Padding; $(\widetilde{G}_i,\widetilde{H}_i)\sim\mathrm{ER}(n,0.5,\rho)$ \\for all $i\in[10]$ with $n\in \{100,500,1000,2000\}$\\ and $r=0.5$.}
\end{subfigure}%
\begin{subfigure}{.5\textwidth}
  \centering
  \includegraphics[width=.7\linewidth]{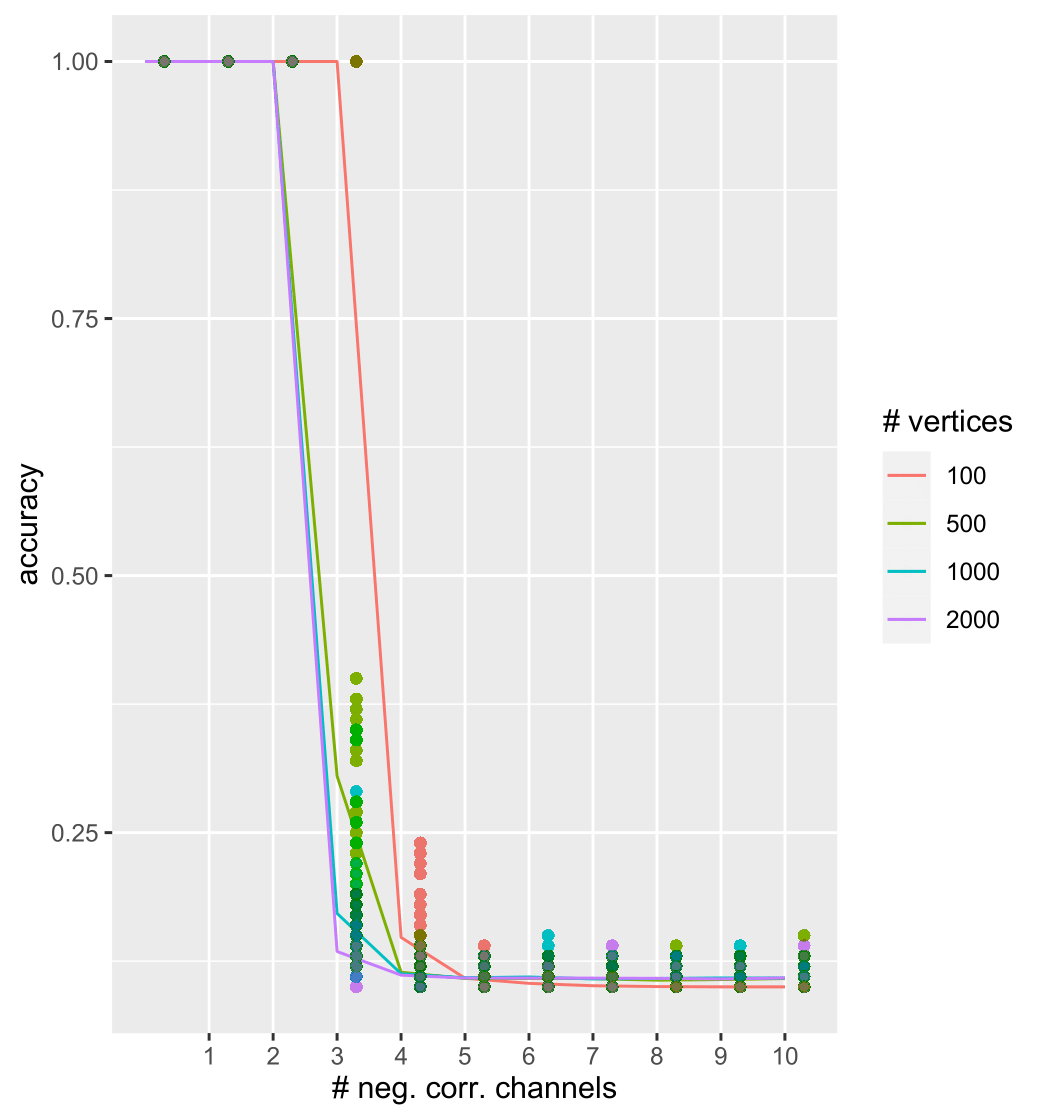}
  \caption{Centered Padding; $(\Hat{G}_i,\Hat{H}_i)\sim\mathrm{ER}(n,0.5,\rho)$ for all $i\in[10]$ with $n\in \{100,500,1000,2000\}$ and $r=0.5$.}
\end{subfigure}
\caption{
Under the MS model setting as in Figure \ref{fig:goodvbad2}, we fix $m=100$ and we apply Naive Padding (left panel) and Centered Padding (right panel) in $({\bf G},{\bf H})\in(\mathcal{M}^{10}_{n},\mathcal{M}^{10}_{100})$ where $n$ varies.
We consider $\rho$ to take two possible values: $\rho=0.5$ for $c_g$ channels, or
$\rho=-0.5$ for $c_b=c-c_g$ channels.
We plot the matching accuracy (averaged over 100 Monte Carlo replicates) obtained by \texttt{M-FAQ} (with 10 seeds) versus $c_b$. In red (resp., green, blue, purple) we plot the results for $n=100$ (resp., $n=500$, $n=1000$, $n=2000$).}
\label{fig:goodvbad2_nnvary}
\end{figure}

Considering $\rho$ to be either $\rho=r$ (for $c_g$ channels) or
$\rho=-r$ (for $c_b=c-c_g$ channels), where $r$ varies in 
$\{0.1,0.2,0.3,0.4,0.5\}$, we plot the matching accuracy (averaged over 2000 (left panel) and 100 (right panel) Monte Carlo replicates) obtained by \texttt{M-FAQ} (with 10 seeds) versus $c_b$ in Figure \ref{fig:goodvbad2}.
For each choice of parameters, we also plot (via the partially transparent points) the accuracy achieved by each MC replicate. 
In red (resp., olive, green, blue, purple) we plot the results for $r=0.1$ (resp., $r=0.2$, $r=0.3$, $r=0.4$, $r=0.5$).
From the figure, we see the expected relationship:  matching at higher levels of $\rho$ yields better accuracy, and more robustness to channels with negative correlation.
Further, we notice that the matching accuracy in the right panel (i.e., $m<n$) is not as good as in the left panel (i.e., $m=n$). We make this phenomenon more clear in Figure \ref{fig:goodvbad2_nnvary}.
\begin{figure}[t!]
\begin{subfigure}{.5\textwidth}
  \centering
  \includegraphics[width=.7\linewidth]{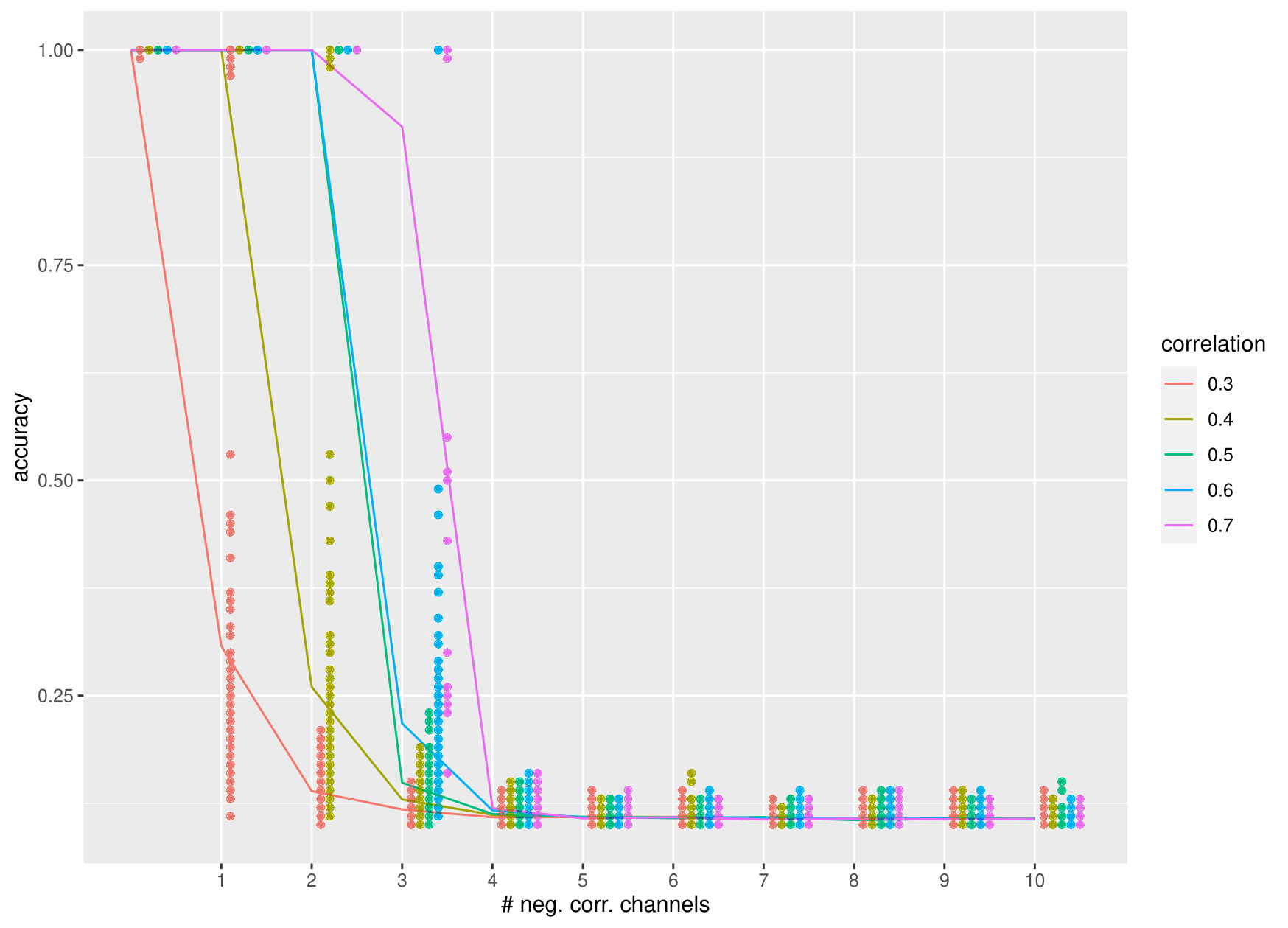}
  \caption{Naive Padding; $(\widetilde{G}_i,\widetilde{H}_i)\sim\mathrm{ER}(500,0.5,\rho)$ for all $i\in[10]$  with $r\in \{0.3,0.4,0.5,0.6,0.7\}$.}
\end{subfigure}
\begin{subfigure}{.5\textwidth}
  \centering
  \includegraphics[width=.7\linewidth]{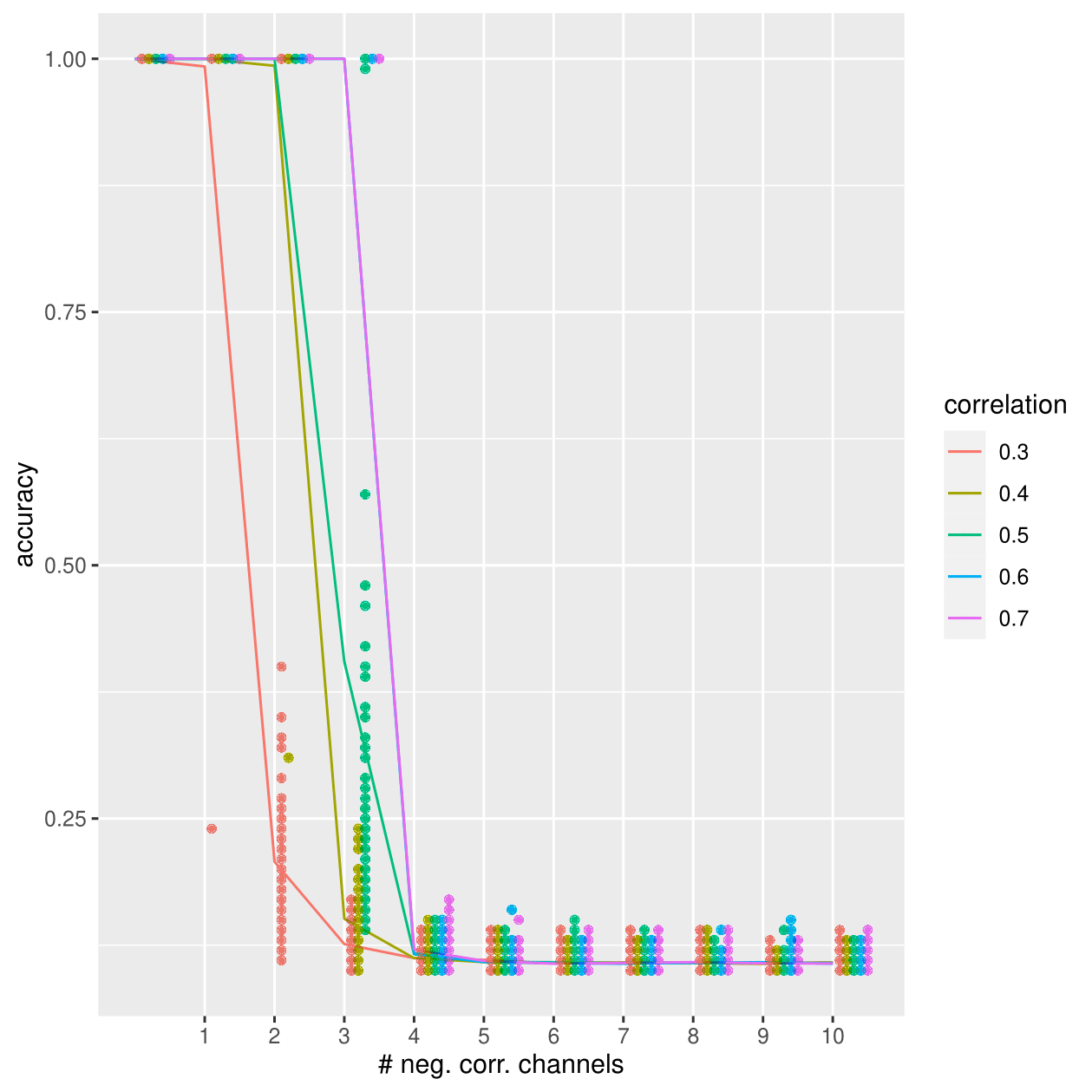}
  \caption{Centered Padding; $(\widehat{G}_i,\widehat{H}_i)\sim\mathrm{ER}(500,0.5,\rho)$ for all $i\in[10]$  with $r\in \{0.3,0.4,0.5,0.6,0.7\}$.}
\end{subfigure}%
\caption{
Under the MS model setting as in Figure \ref{fig:goodvbad2}, we fix $n=500$, $m=100$, and we apply Naive Padding (left panel) and Centered Padding (right panel) in $({\bf G},{\bf H})\in(\mathcal{M}^{10}_{500},\mathcal{M}^{10}_{100})$.
We consider $\rho$ to take two possible values: $\rho=r$ for $c_g$ channels, or
$\rho=-r$ for $c_b=c-c_g$ channels.
We plot the matching accuracy (averaged over 100 Monte Carlo replicates) obtained by \texttt{M-FAQ} (with 10 seeds) versus $c_b$. In red (resp., olive, green, blue, purple) we plot the results for $r=0.3$ (resp., $r=0.4$, $r=0.5$, $r=0.6$, $r=0.7$).}
\label{fig:goodvbad2_rhovary}
\end{figure}

In addition, we study the effect of different padding schemes (Naive vs Centered) in terms of the matching accuracy. We analyze the padding scheme's effectiveness first, by varying the values of the correlation $r\in\{0.3,0.4,0.5,0.6,0.7\}$ while keeping $n,m$ constant (see Figure \ref{fig:goodvbad2_rhovary}) and second, by varying the number of the background vertices $n\in\{100,500,1000,2000\}$ while the template size $m$ remains the same (see Figure \ref{fig:goodvbad2_nnvary}). Using the Naive (resp. Centered) padding scheme, we let $({\bf \widetilde G},{\bf \widetilde H})\in \mathcal{M}^{c}_{n}$ (resp. $({\bf \widehat G},{\bf \widehat H})\in \mathcal{M}^{c}_{n}$) for $c$ ranging over $\{1,2,\cdots,10\}$.
Utilizing $s=10$ seeds, we match ${\bf \widetilde G}$ and ${\bf \widetilde H}$ (resp. ${\bf \widehat G}$ and ${\bf \widehat H}$) using \texttt{M-FAQ} (Algorithm \ref{alg:mfaq}). Results are plotted in Figures \ref{fig:goodvbad2_rhovary} and \ref{fig:goodvbad2_nnvary}. As in Figure \ref{fig:goodvbad2}, we plot the mean matching accuracy (i.e., the fraction of vertices whose latent alignment is recovered correctly) of \texttt{M-FAQ} versus $c_b$, averaged over 100 MC replicates.
For each choice of parameters, we also plot (via the partially transparent points) the accuracy distribution corresponding to the MC replicates. In Figure \ref{fig:goodvbad2_rhovary}, in red (resp., olive, green, blue, purple) we plot the results for $r=0.3$ (resp., $r=0.4$, $r=0.5$, $r=0.6$, $r=0.7$). In Figure \ref{fig:goodvbad2_nnvary}, in red (resp., green, blue, purple) we plot the results for $n=100$ (resp., $n=500$, $n=1000$, $n=2000$).

From Figures \ref{fig:goodvbad2} and \ref{fig:goodvbad2_rhovary}, we observe that  matching at higher levels of $\rho$ yields better accuracy, and more robustness to channels with negative correlation.
Moreover, Figures \ref{fig:goodvbad2_rhovary} and \ref{fig:goodvbad2_nnvary} show that the Centered Padding scheme achieves better performance in terms of matching accuracy than the Naive Padding scheme.


\subsection{ME model matchability}
\label{sec:MEmatch}

To derive analogous results to those in Section \ref{sec:MSmatch} in the ME model, we consider the following setting.
Letting $W\in\gn$ and $T=W[m]$ be the respective background and template source graphs, we again assume that there exist constants $\alpha\leq 1,$ $\beta>0$, and $n_0\in\mathbb{Z}>0$ such that for all $n>n_0$, $m=m(n)$ satisfies $m^\alpha\geq \beta\log n$.
Further assume that for each $i\in[c=c(n)]$ the errorful filters satisfy, 
\begin{align*}
E_i^{(1)}(j,\ell)&=\begin{cases}
s_i=s_i(n) \text{ if }T(j,\ell)=1\\
q_i=q_i(n) \text{ if }T(j,\ell)=0
\end{cases}\\
E_i^{(2)}(j,\ell)&=\begin{cases}
r_i=r_i(n) \text{ if }W(j,\ell)=1\\
t_i=t_i(n) \text{ if }W(j,\ell)=0.
\end{cases}
\end{align*}
For each $P\in\Pi(n)$, define
\begin{align*}
\Delta_P^{(1)}&:=\{ \{j,\ell\}\in\Delta_P\text{ s.t. }T(j,\ell)=1; W(\sigma_p(j),\sigma_p(\ell))=0  \};\\
\Delta_P^{(2)}&:=\{ \{j,\ell\}\in\Delta_P\text{ s.t. }T(j,\ell)=0; W(\sigma_p(j),\sigma_p(\ell))=1  \},
\end{align*}
where $\Delta_P$ is defined as in Eq. (\ref{eq:delp}).
Suppose that there exists an $n_1>0$ such that for all $n>n_1$, we have that for all $k\in[m=m(n)]$ and all $P\in\Pi_{n,m,k}$
\begin{align}
\label{eq:meneed}
|\Delta_P^{(1)}|\sum_i 2(1-2s_i)(1-r_i-t_i)+|\Delta_P^{(2)}|\sum_i 2(1-2q_i)(1-r_i-t_i)\geq k\sqrt{\frac{672 m^{1+\alpha}c}{\beta}},
\end{align}
then
\begin{align}
\label{eq:probME}
\p\left(\argmin_{P\in\Pi_{n}}\sum_{i=1}^{c}\|(\widehat{A}_i\oplus \textbf{0}_{n-m})P-P\widehat{B}_i\|^2_{F}\not\subset \mathcal{P}_{m,n} \right)=2n^{-2};
\end{align}
where the bound in Eq. (\ref{eq:probME}) uses Appendix \ref{app:pfME} and then follows mutatis mutandis from the proof in Appendix \ref{app:MS}.

Exploring this further in the ER setting, consider 
$W\sim ER(n,p=p(n))$ with $p\leq 1/2$.
As in Eq. (\ref{eq:mcd3}), for each $j=1,2$, we then have that 
$$|\Delta_P^{(j)}|\in\left(\,\frac{1}{2}|\Delta_P|p(1-p),\frac{3}{2}|\Delta_P|p(1-p)\,\right),$$ 
with probability at least 
\begin{align}
\label{eq:mcd4}
1-2\text{exp}\left\{\frac{-2 |\Delta_P|p^2(1-p)^2 }{32}\right\}.
\end{align}
Note that if $m>6$, then $mk/3\leq|\Delta_P|\leq mk$, so that with probability at least Eq. (\ref{eq:mcd4}),
$$
|\Delta_P^{(j)}|\in(1/6,3/2)\cdot mkp(1-p).
$$
Suppose that $\alpha<1$, and that there exists an $n_2>0$ such that for all $n>n_2$, we have $mp^2\geq 1344\log n$, and 
\begin{align}
\label{eq:ERneedME}
p\sum_{i=1}^c (1-s_i-q_i)(1-r_i-t_i)\geq\sqrt{\frac{6048m^{\alpha-1}c}{\beta}}.
\end{align}
Then for $n>\max(n_0,n_2)$, $\p(\mathcal{A}_n)\leq 6n^{-2}$.
We have the following theorem (whose proof follows mutatis mutandis to that of Theorem \ref{thm:goodvbad} and so is omitted):
\begin{theorem}
\label{thm:goodvbad2}
With setup as above, 
suppose that $\alpha<1$.
For 
\begin{align*}
c_1=c_1(n) &\text{ channels, suppose that }s_i+q_i=1+e_1>1\text{ and }r_i+t_i=1-e_2<1; \\
c_2=c_2(n) &\text{ channels, suppose that }s_i+q_i=1-e_1<1\text{ and }r_i+t_i=1+e_2>1; \\
c_3=c_3(n) &\text{ channels, suppose that }s_i+q_i=1+e_1>1\text{ and }r_i+t_i=1+e_2>1; \\
c_4=c_4(n) &\text{ channels, suppose that }s_i+q_i=1-e_1<1\text{ and }r_i+t_i=1-e_2<1,
\end{align*}
where $e_1=e_1(n)$ and $e_2=e_2(n)$ can vary in $n$ and $c=c_1+c_2+c_3+c_4$.
Then there exist constants $\gamma,\xi>0$, and $n_2\in\mathbb{Z}>0$ such that 
if for all $n>n_2$
\begin{align}
\label{eq:goodvbadME}
mp^2\geq \xi \log n\notag,\\
pe_1e_2\left[c_3+c_4-c_1-c_2\right]>\gamma\sqrt{m^{\alpha-1}c},
\end{align}
then for $n>\max(n_0,n_2)$, $\p(\mathcal{A}_n)\leq 6n^{-2}.$
If $e_1$, $e_2$, and $c$ are fixed in $n$, we need only require $c_1+c_2<c_3+c_4$ for $\p(\mathcal{A}_n)\leq 6n^{-2}$ to hold for sufficiently large $n$.
\end{theorem}

\section{ Experiments}
\label{sec:exp}

Our previous simulation explored the effect on multiple channels on multiplex matchability.
We next consider the performance of our multiplex matched filter approach in detecting a hidden template in a multilayer social media network from \cite{magnani2011ml}.
The background network contains $3$ aligned channels representing user activity in FriendFeed, Twitter and Youtube (where the Youtube and Twitter channels were generated via FriendFeed which aggregates user information across these platforms).  In total, there are $6,407$ unique vertices across the three channels, with the channel specific networks satisfying:
\begin{center}
\begin{tabular}{c|c|c}
channel& vertices & edges\\
\hline
FriendFeed& 5,540 &  31,921 \\
Twitter& 5,702 &  42,327 \\
YouTube& 663 &  614 
\end{tabular}
\end{center}
Given a 35 vertex multiplex template ${\bf H}$ created by  
Pacific Northwest National Laboratories for the DARPA MAA program, we ran our \texttt{M-GMMF} algorithm (Algorithm \ref{alg:mgmmf}) to attempt to recover the template in ${\bf G}$; results are summarized below.

\begin{figure}[t!]
\centering
\begin{tabular}{cc}
\includegraphics[width=0.5\textwidth]{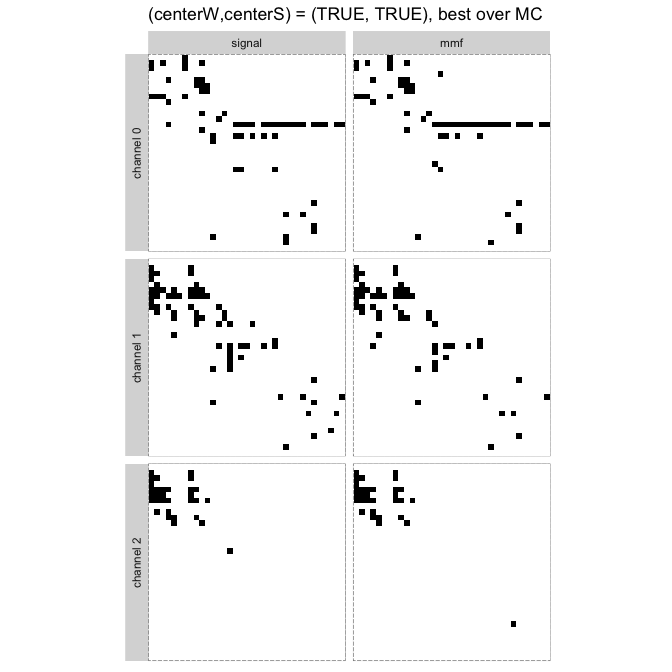}&\hspace{-15mm}\includegraphics[width=0.5\textwidth]{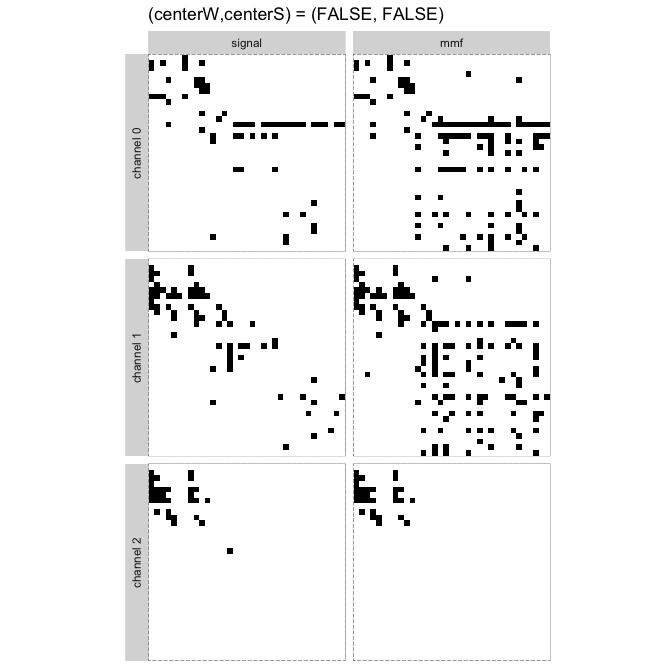}\\
\end{tabular}
\caption{Signal recovered by the best performing random restart in \texttt{M-GMMF} across different Padding regimes. 
As in Table \ref{tab:young2}, the best performer is the one that recovers the highest average $\%$ of the template edges across the three channels (averaging the $\%$ within each channel across channels).
In the left panel, we plot the Centered Padding regime and in the right panel the Naive Padding regime.
For each centering regime, we plot the signal template across the three channels (in the left 3 panels) and the best recovered subgraphs in the background (in the right 3 panels).}
  \label{fig:o1}
\end{figure}

In our first experiment, we first considered running ``cold-start'' \texttt{M-GMMF}; that is, no prior information (in the form of seeds, hard or soft) is utilized in the algorithm.
We consider padding the graph via the Naive Padding and Centered Padding regimes of Section \ref{sec:pad}, and for each padding regime, we ran \texttt{M-GMMF} with $N=100$ random restarts.
Numeric results are summarized in Table \ref{tab:young2} (with the best recovered background signals also plotted in Figure \ref{fig:o1}).
While the best recovered signal in the Naive Padding regime captures all but two template edges, this is at the expense of many extraneous background edges that do not appear in the template.
On the other hand, the Centered Padding regime recovers most of the template edges (across the three channels) with minimal extra template edges in the recovered signal.

\begin{table}[t!]
 \centering
 \begin{tabular}{c|c|c|c}
Padding regime & $\%$ recovered in ch. 1& $\%$ recovered in ch. 2& $\%$ recovered in ch. 3\\
\hline
Centered& 86.67&  85.07& 96.77\\
Naive & 98.33&  100&      96.77\\
\end{tabular}
\caption{
For each padding regime, we provide the $\%$ of template edges present in the recovered background signal in the best random restart. 
For example, the best recovered background signal in the Centered Padding regime recovered $86.67\%$ of the edges in template channel 1, and $85.07\%$ of the edges in template channel 2, and $96.77\%$ of the edges in template channel 3.
Here, the best performer is the one that recovers the highest average $\%$ across the three channels (averaging the $\%$ within each channel across channels).}
\label{tab:young2}
 \end{table}

The \texttt{M-FAQ} algorithmic primitive (Algorithm \ref{alg:mfaq}) used in our implementation of \texttt{M-GMMF} is most effective when it can leverage a priori available matching data in the form of \emph{seeded} vertices.
Seeds can either come in the form of {\it hard seeds} (a priori known 1--to--1 matches; here that would translate to template vertices whose exact match is known in the background) or {\it soft seeds} (where a soft seeded vertex $v$ in ${\bf H}$ has an a priori known distribution over possible matches in ${\bf G}$; 
here this would translate into template vertices with a list of candidate matches in the background). 
While hard seeds are costly and often unavailable in practice, there are many scalable procedures in the literature for automatically generating soft seed matches.
Here, we use as a soft-seeding the output of \cite{moorman2018filtering,moorman2021subgraph}, a filtering approach for finding all subgraphs of the background network homomorphic to the template.

For each node in the template, the output of \cite{moorman2018filtering,moorman2021subgraph} produces a multiset of candidate matches in the background, where each candidate match corresponds to a template copy contained in the background as a subgraph (not necessarily as an induced subgraph).
We convert the candidate matches into probabilities by simply converting the multiset to a count vector and normalizing the count vector to sum to $1$.
We then consider the normalized count vectors as rows of a stochastic matrix; this stochastic matrix provides \texttt{M-FAQ} with a soft-seeding which can be used to initialize the algorithm.

Considering random restarts as perturbations (akin to  Step 2 of Algorithm \ref{alg:mgmmf}) of the soft-seeding (conditioned on retaining nonnegative entries), we ran \texttt{M-GMMF} using a generalization of the Centered Padding regime, which is defined as follows:
For each $i\in[c]$, define the weighted adjacency matrices $\breve{A}_i\in \R^{m\times m}$ and $\breve{B}_i\in \R^{n\times n}$ via 
\begin{align}
\label{eq:cenpadmodified}
 \breve{A}_i(u,v)&=
    \begin{dcases}
        1 & \text{if u,v}\in V(H_i), \text{ and \{u,v\}}\in E(H_i); \\
        -w & \text{if u,v}\in V(H_i), \text{ and \{u,v\}}\notin E(H_i); \\
        0 & \text{if u or v}\in [m]\setminus V(H_i); \\
    \end{dcases}\\
 \breve{B}_i(u,v)&=
    \begin{dcases}
        1 & \text{if u,v}\in V(G_i), \text{ and \{u,v\}}\in E(G_i); \\
        -1 & \text{if u,v}\in V(G_i), \text{ and \{u,v\}}\notin E(G_i); \\
        0 & \text{if u or v}\in [n]\setminus V(G_i); \\
    \end{dcases}\notag
\end{align}
where we vary $w$ from $0$ to $1$.
Note that $w=0$ yields Naive Padding, and $w=1$ yields Centered Padding.
Optimal performance in the present experiment was achieved with $w=0.25$, in which case $N=4000$ random restarts yielded an induced subgraph in the background that was isomorphic to the template network.

\subsection{M-GMMF on Semantic Property Graphs}
\label{sec:find}
For our second example, we consider  the semantic property graph released by Pacific Northwest National Laboratories as part of the MAA-AIDA Data Release V2.1.2 via the DARPA MAA program \cite{ebsch2020using}. 
In this dataset, the background network is a knowledge graph constructed from a variety of documents (e.g., newspaper articles) by DARPA's AIDA program.
At a high level, the graph is encoding the real-world relationships between a variety of entities (people, locations, major events, etc.) that can be automatically extracted from a variety of data sources.
Practically, the graph is a richly featured network on the order of 100K nodes.
Node properties include name; rdf:type (corresponding to a structured ontology of types); textValue; linkTarget; start time; among others.
Edge properties include name/id, rdf:type,  argument (values given to edges of a given rdf:type), among others.
Note that many nodes and edges do not have values for all properties.
The templates here, themselves richly featured knowledge graphs, are of the order of 10s of vertices (ranging in size from 33 nodes/40 edges to 11 nodes/11 edges); for each of three template types, there are 6 variants with varying error levels, including one variant (version ``A'') that is perfectly/isomorphically embedded into the background.
\begin{figure}[t!]
\centering
\includegraphics[width=0.8\textwidth]{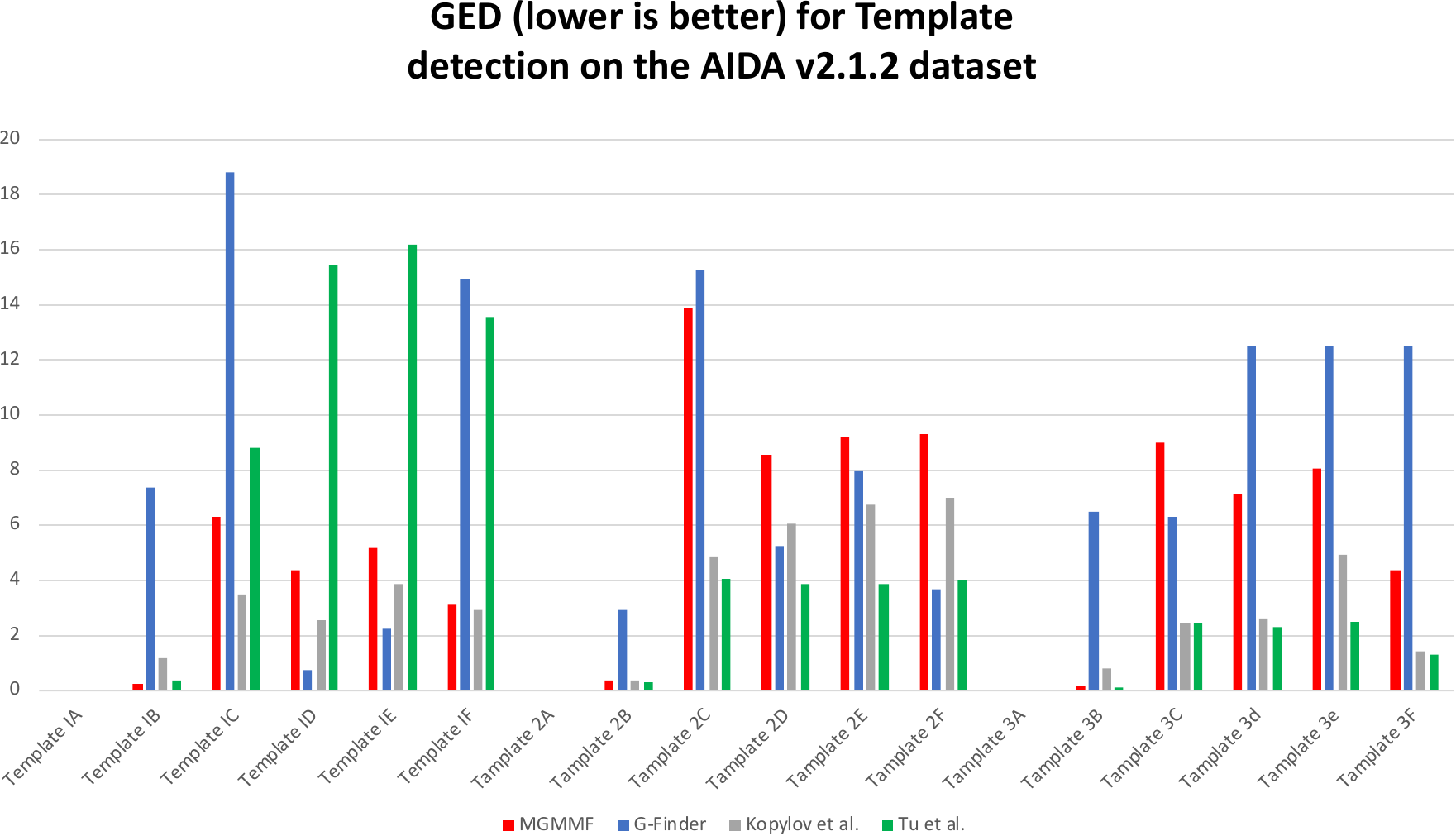}
\caption{Graph edit distance score from \cite{ebsch2020using} for the best recovered signal for the filter algorithms of \cite{tu2020inexact} (green) and \cite{kopylov2020semantic} (gray); G-finder of \cite{gfinder} (blue; as implemented in \cite{kopylov2020semantic}) and M-GMMF (red; with 1000 random restarts) for each of the 18 templates.
}
  \label{fig:met}
\end{figure}

The principle challenge in applying our \texttt{M-GMMF} methodology on such richly featured data is sensibly incorporating the rich, structured features into our multiplex network framework.
Towards this end, we adopted the following approach. 
We incorporated the vertex features/properties into a penalty term in the objective function, encoding the features into a vertex--to--vertex similarity matrix $S$ (this is possible provided that similarities are easily computed within each vertex covariate, which is the case here).
Edge features were used to divide the knowledge graph into multiple overlapping channels in a multiplex network.
One channel was assigned to each unique (\texttt{E(rdf:type)}, \texttt{E(argument)}) pair in the template, and we divided background edges amongst the channels via
\begin{itemize}
\item[i.] A hard split based on \texttt{E(argument)}: Within each (\texttt{E(rdf:type)}, \texttt{E(argument)}) channel, only edges with matching \texttt{E(argument)} are potentially present.  If the \texttt{E(argument)} property is missing in the background, we allow the edge to possibly exist in all (\texttt{E(rdf:type)}, \texttt{E(argument)}) channels.
\item[ii.] A soft split (weighted according to \texttt{E(rdf:type)} similarity; note that a \texttt{rdf:type} similarity function was provided with the data) based on \texttt{E(rdf:type)}:  Within each (\texttt{E(rdf:type)}, \texttt{E(argument)}) channel, each background edge with matching (or missing) \texttt{E(argument)} is present, and the edge is weighted according to a similarity measured between its \texttt{E(rdf:type)} and that of the channel. Scalability gains can be achieved by thresholding the similarities to improve sparsity.
\end{itemize}
Spatiotemporal constraints can be coded into separate channels in the multiplex graph, one channel per spatiotemporal constraint in the template.
Each constraint (e.g., action A must occur between x and y days after action B) yields an edge filter, with only edges that could potentially satisfy the constraint being added to that constraint channel.
Lastly, numeric edge features are used to further weight the edges in the background/template.
The final objective function we used in our \texttt{M-GMMF} approach was then of the form
\begin{equation}
\label{eq:cmgmp2}
\argmax_{P\in\Pi_{n}}\sum_{i=1}^{c}\text{tr}((\widehat{A}_i\oplus \textbf{0}_{n-m})P\widehat{B}_iP^T)+\text{tr}(SP^T).
\end{equation}
Performance of M-GMMF and various other approaches (as scored by the GED scoring metric of \cite{ebsch2020using}) are presented in Figure \ref{fig:met} (Note that the filtering approach of Tu et al., as presented in \cite{tu2020inexact}, also present a method that uses the clean ``A'' template for training, and achieves essentially the best score for all template versions; we did not include those scores for comparison in our figure). 

\begin{figure}[tbh!]
\centering
\includegraphics[width=1\textwidth]{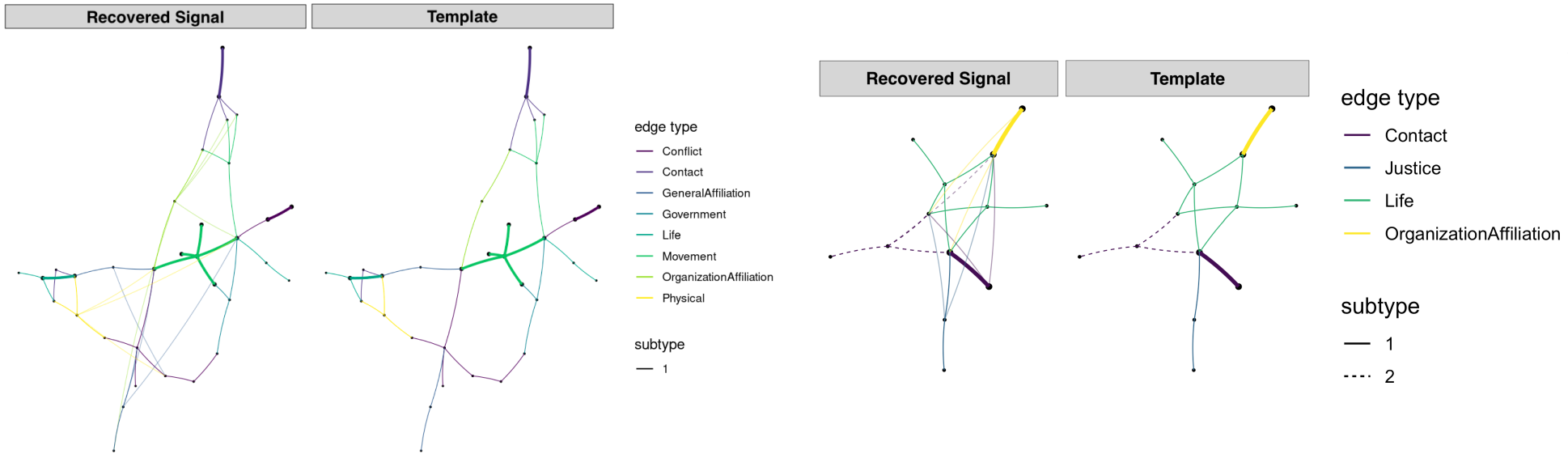}
\caption{Example of template knowledge graph and recovered signal in the knowledge graph.
In each graph, different colored edges represent different \texttt{E(rdf:types)}, with the dotted versus solid edges representing different \texttt{E(arguments)}.}
  \label{fig:ma2}
\end{figure}
Overall, 18 templates were present: 3 distinct template types, each with 6 variants encompassing different amounts of template noise which give an indication of the knowledge graph structure encoded into the multiplex graph; see Figure \ref{fig:ma2}, for a pair of example templates and M-GMMF recovered signals here.
One template of each type was constructed to have a perfect isomorphic match in the background, while the noisy templates were designed for inexact/fuzzy matching.
All approaches identified the isomorphic match in the background for all three template types version ``A'', while our approach achieved its best relative results on the larger, more complex template (Template 1). 
For Template 1, we obtained the best (or nearly the best) score on 3/6 versions; see Table \ref{tab:T1} for detail.
\begin{table}[htb!]
\center
\begin{tabular}{c||c|c|c|c|c|c}
Method & T1A & T1B & T1C & T1D& T1E & T1F\\
\hline\hline
M-GMMF & 0 &0.24 &6.28 &4.35 &5.17 &3.12\\
\hline
G-Finder & 0 & 7.38 & 18.81 & 0.76 & 2.25 & 14.96\\
\hline
Tu et al. \cite{tu2020inexact} & 0 & 0.347 & 8.783 & 15.47 & 19.194 & 13.596\\
\hline
Kopylov et al. \cite{kopylov2020semantic} & 0 & 1.17 & 3.47 & 2.58 & 3.85 & 2.94\\
\end{tabular}
\caption{Performance on Template 1}
\label{tab:T1}
\end{table}
Templates 2 and 3 were essentially tree-like structures (nodes/edges is 13/15 and 11/11 respectively), and we suspect that the filtering-based approaches are more suitable to this problem type.
Indeed, M-GMMF is designed for larger/more complex templates, though our performance (especially compared to the non-filtering G-Finder of \cite{gfinder}) is encouraging on these instantiations, especially on version ``B'' of the templates.

\begin{figure}[t!]
\centering
\includegraphics[width=1\textwidth]{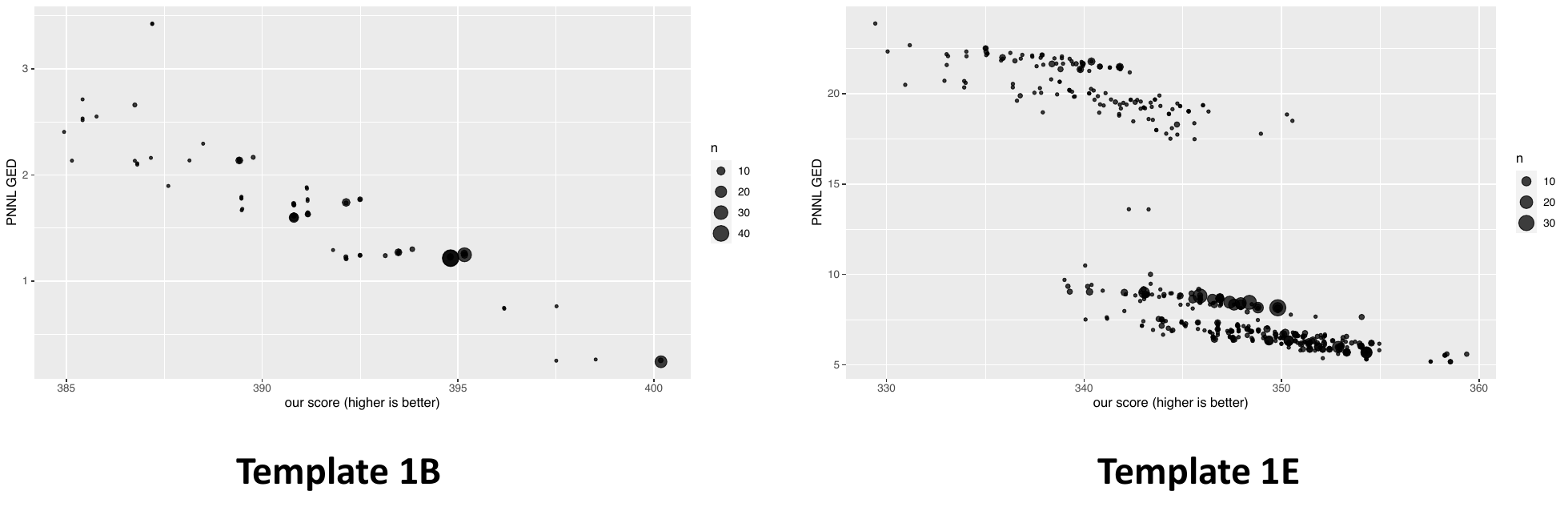}
\caption{For two of the templates, we plot the scores of the recovered signals (recovered by M-GMMF across multiple random restarts), where scores are computed both using the GED metric of \cite{ebsch2020using} (y-axis) and the GM objective function (x-axis).}
  \label{fig:ged2}
\end{figure}

Note also that our approach does not directly seek to optimize the GED of \cite{ebsch2020using} and here does not make use of the importance weights provided by the GED scoring metric (though these could easily be encoded via edge weights and scaling the similarities in $S$); rather, we seek to optimize the multiplex GM objective function of \ref{eq:cmgmp2}.
Nonetheless, as shown in Figure \ref{fig:ged2}, the rankings of the random restart outputs for our GM objective function and for the GED are often highly correlated, with recovered signals scoring well in one metric often scoring well in both.
The interesting gap appearing in the plot of Template 1E is evocative of the GM phase transitions appearing in the literature (see, for example, \cite{fishkind2013consistent}) and bears further study.



\section{Discussion}
\label{sec:discussion}

In this paper, we presented a framework for finding noisy copies of a multiplex template in a large multiplex background network.
Our strategy, which extends \cite{sussman2018matched}  to the multiplex setting, uses graph matching combined with multiple random restarts to search the background for locally optimal matches to the template.
To formalize this strategy, we provided a very natural extension of the classical graph matching problem to the multiplex that is easily amended to matching graphs of different orders (both across networks and channels).
Further, the effectiveness of the resulting algorithm, named \texttt{M-GMMF}, is demonstrated both theoretically and empirically.

There are a number of extensions and open questions that arose during the course of this work.
Natural theoretic extensions include lifting Theorems \ref{thm:goodvbad} and \ref{thm:goodvbad2} to non-edge independent models (note that certain localized dependencies amongst edges can easily be handled in the McDiarmind proof framework, while globally dependent errors provide a more significant challenge); 
formulating the analogues of Theorems \ref{thm:goodvbad} and \ref{thm:goodvbad2} in the weighted, attributed graph settings; 
and considering the theoretic properties of various continuous relaxations of the multiplex GM problem akin to \cite{aflalo2015convex,lyz2016relax,bento2018family}.

A key methodological questions in multiplex graph matching was touched upon in Remark \ref{app:rem1}; indeed, we expect the question of how to weight the matching across channels to be essential when applying these methods to topologically diverse and weighted networks.
If the order of magnitude of edge weights vary across channel, then it is easy to see a GM algorithm aligning channels with large edge weights at the expense of the alignment accuracy in other channels.
Judiciously choosing $(\lambda_i)$ would allow for the signal in channels with smaller edge weights to be better leveraged towards a better overall matching.

While the largest network we consider in this work has $\approx 100,000$ vertices, scaling this approach to very large networks is essential.
By utilizing efficient data structures for sparse, low-rank matrices and a clever implementation of the LAP subroutine of $\texttt{M-FAQ}$ (step iii. in Algorithm \ref{alg:mfaq}), we are able to match $O(10)$ vertex templates to $20K$-vertex background graphs in $< 10s$ per restart with our base \texttt{M-GMMF} code (available in \texttt{iGraphMatch}) implemented in \texttt{R} on a standard laptop.
Further work to scale \texttt{M-GMMF} by leveraging both efficient data structures and scalable approximate LAP solvers is currently underway.
\subsection*{List of abbreviations}
\begin{acronym}[MPC] 
\acro{GM(P)}{Graph Matching (Problem)}
\acro{nMGMP}{Naive Multiplex Graph Matching Problem}
\acro{cMGMP}{Centered Multiplex Graph Matching Problem}
\acro{M-GMMF}{Multiplex Graph Matching Matched Filters}
\acro{M-FAQ}{Multiplex Fast Approximate Quadratic}
\acro{ME}{(Single Channel Source) Error Multiplex}
\acro{MS}{(Single Channel Errors) Source Multiplex}
\acro{GED}{Graph Edit Distance}
\acro{LAP}{Linear Assignment Program}
\acro{DARPA}{Defense Advanced Research Projects Agency}
\acro{MAA}{Modeling Adversarial Activity}
\acro{AIDA}{Active Interpretation of Disparate Alternatives}
\end{acronym}
\section*{Declarations}
\subsection*{Availability of data and materials}
 The background graphs for the 3-channel social network in Section  \ref{sec:exp} is available at \url{http://multilayer.it.uu.se/datasets.html}. 
 The data from the DARPA MAA analysis Section is not publicly available, and the obtained results for the outside algorithms appear in the cited papers in the literature. 
 
 \noindent
The code implementing the \texttt{M-GMMF} and \texttt{M-FAQ} procedures can be downloaded as part of our \texttt{R} package, \texttt{iGraphMatch}, which is available on \texttt{CRAN} or can be downloaded at \url{https://github.com/dpmcsuss/iGraphMatch}.

\subsection*{Competing interests}
Not applicable.
\subsection*{Funding}

Dr. Sussman's contribution to this work was partially supported by a grant from MIT Lincoln Labs and the Department of Defense.
This material is based on research sponsored by the Air Force Research Laboratory and DARPA under agreement numbers FA8750-18-2-0035 and FA8750-20-2-1001.
The U.S.
Government is authorized to reproduce and distribute reprints for Governmental purposes notwithstanding any copyright notation thereon. The views
and conclusions contained herein are those of the authors and should not
be interpreted as necessarily representing the official policies or endorsements, either expressed or implied, of the Air Force Research Laboratory
and DARPA or the U.S. Government.

\subsection*{Authors' contributions}
Daniel L. Sussman, Carey E. Priebe and Vince Lyzinski, and Konstantinos Pantazis conceived of the method and developed the theory.
Konstantinos Pantazis, Youngser Park, Daniel L. Sussman, Vince Lyzinski and Zhirui Li worked on writing and implementing the algorithm and performing relevant experiments. Konstantinos Pantazis, Daniel L. Sussman and Vince Lyzinski wrote and curated the manuscript.

\subsection*{Acknowledgments}
Not applicable.


\bibliographystyle{plain}
\bibliography{biblio}

\begin{thebibliography}{10}

\bibitem{aflalo2015convex}
Y.~Aflalo, A.~Bronstein, and R.~Kimmel.
\newblock On convex relaxation of graph isomorphism.
\newblock {\em Proceedings of the National Academy of Sciences},
  112(10):2942--2947, 2015.

\bibitem{colorcoding2}
N.~Alon, P.~Dao, I.~Hajirasouliha, F.~Hormozdiari, and S.~C. Sahinalp.
\newblock Biomolecular network motif counting and discovery by color coding.
\newblock {\em Bioinformatics}, 24(13):i241--i249, 2008.

\bibitem{colorcoding1}
N.~Alon, R.~Yuster, and U.~Zwick.
\newblock Color-coding.
\newblock {\em Journal of the ACM (JACM)}, 42(4):844--856, 1995.

\bibitem{arroyo2018maximum}
J.~Arroyo, D.~L. Sussman, C.~E. Priebe, and V.~Lyzinski.
\newblock Maximum likelihood estimation and graph matching in errorfully
  observed networks.
\newblock {\em arXiv preprint arXiv:1812.10519}, 2018.

\bibitem{ma9}
B.~Barak, C.~Chou, Z.~Lei, T.~Schramm, and Y.~Sheng.
\newblock (nearly) efficient algorithms for the graph matching problem on
  correlated random graphs.
\newblock {\em arXiv preprint arXiv:1805.02349}, 2018.

\bibitem{bento2018family}
J.~Bento and S.~Ioannidis.
\newblock A family of tractable graph distances.
\newblock {\em arXiv preprint arXiv:1801.04301}, 2018.

\bibitem{boccaletti2014structure}
S.~Boccaletti, G.~Bianconi, R.~Criado, C.~I. Del~Genio, J.~G{\'o}mez-Gardenes,
  M.~Romance, I.~Sendina-Nadal, Z.~Wang, and M.~Zanin.
\newblock The structure and dynamics of multilayer networks.
\newblock {\em Physics Reports}, 544(1):1--122, 2014.

\bibitem{jointLi}
L.~Chen, J.~T. Vogelstein, V.~Lyzinski, and C.~E. Priebe.
\newblock A joint graph inference case study: {T}he {C}. elegans chemical and
  electrical connectomes.
\newblock {\em Worm}, 5(2), 2016.

\bibitem{ConteReview}
D.~Conte, P.~Foggia, C.~Sansone, and M.~Vento.
\newblock Thirty years of graph matching in pattern recognition.
\newblock {\em International Journal of Pattern Recognition and Artificial
  Intelligence}, 18(03):265--298, 2004.

\bibitem{feasibilityrules}
L.~P. Cordella, P.~Foggia, C.~Sansone, and M.~Vento.
\newblock A (sub) graph isomorphism algorithm for matching large graphs.
\newblock {\em IEEE Transactions on Pattern Analysis and Machine Intelligence},
  26(10):1367--1372, 2004.

\bibitem{ma8}
D.~Cullina and N.~Kiyavash.
\newblock Improved achievability and converse bounds for erdos-r{\'e}nyi graph
  matching.
\newblock In {\em ACM SIGMETRICS Performance Evaluation Review}, volume~44,
  pages 63--72. ACM, 2016.

\bibitem{ma7}
D.~Cullina and N.~Kiyavash.
\newblock Exact alignment recovery for correlated erdos-renyi graphs.
\newblock {\em arXiv preprint arXiv:1711.06783}, 2017.

\bibitem{ma10}
D.~Cullina, N.~Kiyavash, P.~Mittal, and H.~V. Poor.
\newblock Partial recovery of {E}rdos-{R}enyi graph alignment via $ k $-core
  alignment.
\newblock {\em arXiv preprint arXiv:1809.03553}, 2018.

\bibitem{ma1}
J.~Ding, Z.~Ma, Y.~Wu, and J.~Xu.
\newblock Efficient random graph matching via degree profiles.
\newblock {\em arXiv preprint arXiv:1811.07821}, 2018.

\bibitem{ebsch2020using}
C.~L. Ebsch, J.~A. Cottam, N.~C. Heller, R.~D. Deshmukh, and G.~Chin.
\newblock Using graph edit distance for noisy subgraph matching of semantic
  property graphs.
\newblock In {\em 2020 IEEE International Conference on Big Data (Big Data)},
  pages 2520--2525. IEEE, 2020.

\bibitem{gmrev}
F.~Emmert-Streib, M.~Dehmer, and Y.~Shi.
\newblock Fifty years of graph matching, network alignment and network
  comparison.
\newblock {\em Information sciences}, 346--347:180--197, 2016.

\bibitem{fishkind2013consistent}
D.~E. Fishkind, D.~L. Sussman, M.~Tang, J.~T. Vogelstein, and Carey~E Priebe.
\newblock Consistent adjacency-spectral partitioning for the stochastic block
  model when the model parameters are unknown.
\newblock {\em SIAM Journal on Matrix Analysis and Applications}, 34:23--39,
  2013.

\bibitem{ModFAQ}
D.E. Fishkind, S.~Adali, H.G. Patsolic, L.~Meng, D.~Singh, V.~Lyzinski, and
  C.E. Priebe.
\newblock Seeded graph matching.
\newblock {\em Pattern Recognition}, 87:203 -- 215, 2019.

\bibitem{foggia2014graph}
P.~Foggia, G.~Percannella, and M.~Vento.
\newblock Graph matching and learning in pattern recognition in the last 10
  years.
\newblock {\em International Journal of Pattern Recognition and Artificial
  Intelligence}, 28(01):1450001, 2014.

\bibitem{graphhomobased}
F.~V. Fomin, D.~Lokshtanov, V.~Raman, S.~Saurabh, and B.~V.~R. Rao.
\newblock Faster algorithms for finding and counting subgraphs.
\newblock {\em Journal of Computer and System Sciences}, 78(3):698--706, 2012.

\bibitem{goga2015reliability}
O.~Goga, P.~Loiseau, R.~Sommer, R.~Teixeira, and K.~P. Gummadi.
\newblock On the reliability of profile matching across large online social
  networks.
\newblock In {\em Proceedings of the 21th ACM SIGKDD International Conference
  on Knowledge Discovery and Data Mining}, pages 1799--1808. ACM, 2015.

\bibitem{ma2}
E.~Kazemi, S.~H. Hassani, and M.~Grossglauser.
\newblock Growing a graph matching from a handful of seeds.
\newblock {\em Proceedings of the VLDB Endowment}, 8(10):1010--1021, 2015.

\bibitem{ma6}
E.~Kazemi, L.~Yartseva, and M.~Grossglauser.
\newblock When can two unlabeled networks be aligned under partial overlap?
\newblock In {\em Communication, Control, and Computing (Allerton), 2015 53rd
  Annual Allerton Conference on}, pages 33--42. IEEE, 2015.

\bibitem{kivela2014multilayer}
M.~Kivel{\"a}, A.~Arenas, M.~Barthelemy, J.~P. Gleeson, Y.~Moreno, and M.~A.
  Porter.
\newblock Multilayer networks.
\newblock {\em Journal of complex networks}, 2(3):203--271, 2014.

\bibitem{kivela2017isomorphisms}
M.~Kivel{\"a} and M.~A. Porter.
\newblock Isomorphisms in multilayer networks.
\newblock {\em IEEE Transactions on Network Science and Engineering}, 2017.

\bibitem{kopylov2020semantic}
A.~Kopylov, J.~Xu, K.~Ni, S.~Roach, and T.-C. Lu.
\newblock Semantic guided filtering strategy for best-effort subgraph matching
  in knowledge graphs.
\newblock In {\em 2020 IEEE International Conference on Big Data (Big Data)},
  pages 2539--2545. IEEE, 2020.

\bibitem{sgsurvey}
J.~Lee, W.~Han, R.~Kasperovics, and J.~Lee.
\newblock An in-depth comparison of subgraph isomorphism algorithms in graph
  databases.
\newblock In {\em Proceedings of the VLDB Endowment}, volume~6, pages 133--144.
  VLDB Endowment, 2012.

\bibitem{gfinder}
L.~Liu, B.~Du, and H.~Tong.
\newblock G-finder: Approximate attributed subgraph matching.
\newblock In {\em 2019 IEEE International Conference on Big Data (Big Data)},
  pages 513--522. IEEE, 2019.

\bibitem{lyz2016relax}
V.~Lyzinski, D.~E. Fishkind, M.~Fiori, J.~T. Vogelstein, C.~E. Priebe, and
  G.~Sapiro.
\newblock Relax at your own risk.
\newblock {\em IEEE Transactions on Pattern Analysis and Machine Intelligence},
  pages 60--73, 2016.

\bibitem{ma5}
V.~Lyzinski and D.~L. Sussman.
\newblock Matchability of heterogeneous networks pairs.
\newblock {\em arXiv preprint arXiv:1705.02294}, 2017.

\bibitem{magnani2011ml}
M.~Magnani and L.~Rossi.
\newblock The ml-model for multi-layer social networks.
\newblock In {\em 2011 International Conference on Advances in Social Networks
  Analysis and Mining}, pages 5--12. IEEE, 2011.

\bibitem{mcdiarmid1989method}
C.~McDiarmid.
\newblock On the method of bounded differences.
\newblock {\em Surveys in combinatorics}, 141(1):148--188, 1989.

\bibitem{moorman2018filtering}
J.~D. Moorman, Q.~Chen, T.~K. Tu, Z.~M. Boyd, and A.~L. Bertozzi.
\newblock Filtering methods for subgraph matching on multiplex networks.
\newblock In {\em 2018 IEEE International Conference on Big Data (Big Data)},
  pages 3980--3985. IEEE, 2018.

\bibitem{moorman2021subgraph}
J.~D. Moorman, T.~Tu, Q.~Chen, X.~He, and A.~Bertozzi.
\newblock Subgraph matching on multiplex networks.
\newblock {\em IEEE Transactions on Network Science and Engineering}, 2021.

\bibitem{mucha2010community}
P.~J. Mucha, T.~Richardson, K.~Macon, M.~A. Porter, and J.~Onnela.
\newblock Community structure in time-dependent, multiscale, and multiplex
  networks.
\newblock {\em Science}, 328(5980):876--878, 2010.

\bibitem{ng2011multirank}
M.~K. Ng, X.~Li, and Y.~Ye.
\newblock Multirank: co-ranking for objects and relations in multi-relational
  data.
\newblock In {\em Proceedings of the 17th ACM SIGKDD international conference
  on Knowledge discovery and data mining}, pages 1217--1225. ACM, 2011.

\bibitem{ma4}
E.~Onaran, S.~Garg, and E.~Erkip.
\newblock Optimal de-anonymization in random graphs with community structure.
\newblock In {\em 2016 50th Asilomar Conference on Signals, Systems and
  Computers}, pages 709--713. IEEE, 2016.

\bibitem{ma3}
P.~Pedarsani and M.~Grossglauser.
\newblock On the privacy of anonymized networks.
\newblock In {\em Proceedings of the 17th ACM SIGKDD international conference
  on Knowledge discovery and data mining}, pages 1235--1243. ACM, 2011.

\bibitem{priebe2015statistical}
C.~E. Priebe, D.~L. Sussman, M.~Tang, and J.~T. Vogelstein.
\newblock Statistical inference on errorfully observed graphs.
\newblock {\em Journal of Computational and Graphical Statistics},
  24(4):930--953, 2015.

\bibitem{read1977graph}
R.~C. Read and D.~G. Corneil.
\newblock The graph isomorphism disease.
\newblock {\em Journal of Graph Theory}, 1(4):339--363, 1977.

\bibitem{sussman2018matched}
D.~L. Sussman, V.~Lyzinski, Y.~Park, and C.~E. Priebe.
\newblock Matched filters for noisy induced subgraph detection.
\newblock {\em Pattern Analysis and Machine Intelligence, IEEE Transactions
  on}, accepted for publication.

\bibitem{takes2017detecting}
F.~W. Takes, W.~A. Kosters, and B.~Witte.
\newblock Detecting motifs in multiplex corporate networks.
\newblock In {\em International Workshop on Complex Networks and their
  Applications}, pages 502--515. Springer, 2017.

\bibitem{tu2020inexact}
T.~K. Tu, J.~D. Moorman, D.~Yang, Q.~Chen, and A.~L. Bertozzi.
\newblock Inexact attributed subgraph matching.
\newblock In {\em 2020 IEEE International Conference on Big Data (Big Data)},
  pages 2575--2582. IEEE, 2020.

\bibitem{efficienttreesearch}
J.~R. Ullmann.
\newblock An algorithm for subgraph isomorphism.
\newblock {\em Journal of the ACM (JACM)}, 23(1):31--42, 1976.

\bibitem{white1986structure}
J.~G. White, E.~Southgate, J.~N. Thomson, and S.~Brenner.
\newblock The structure of the nervous system of the nematode caenorhabditis
  elegans.
\newblock {\em Philos Trans R Soc Lond B Biol Sci}, 314(1165):1--340, 1986.

\bibitem{yan2016short}
J.~Yan, X.~Yin, W.~Lin, C.~Deng, H.~Zha, and X.~Yang.
\newblock A short survey of recent advances in graph matching.
\newblock In {\em Proceedings of the 2016 ACM on International Conference on
  Multimedia Retrieval}, pages 167--174. ACM, 2016.

\bibitem{yang2016mining}
B.~Yang and J.~Liu.
\newblock Mining multiplex structural patterns from complex networks.
\newblock In {\em Wisdom Web of Things}, pages 275--301. Springer, 2016.

\end{thebibliography}

\appendix
\section{Appendix}
Herein we collect details of our auxiliary algorithms and proofs of our main results.

\subsection{Multiplex FAQ}
\label{app:mfaq}
The details of the \texttt{M-FAQ} algorithm are presented below.

\begin{algorithm}[h!]
  \begin{algorithmic}
    \STATE \textbf{Input}: Multiplex graphs $\bf{A}\in\mathcal{M}_m^c$ and $\bf{B}\in\mathcal{M}_n^c$; weights $(\lambda_i)$; padding regime; tolerance $\epsilon\in\mathbb{R}>0$; initialization $P^{(0)}$
\vspace{3mm}

\STATE Pad $\bf{A}$ and $\bf{B}$ accordingly; Denote the modified, padded multiplex graphs via $\bf{H}\,(=\tba\text{ or }\widehat A\,)$ and $\bf{G}\,(=\widetilde B\text{ or }\widehat B\,)$
\WHILE {$\|P^{(t)}-P^{(t-1)}\|_F>\epsilon$}
\STATE{\bf i.  } $P^{(t)}\leftarrow P^{(t-1)}$
\STATE{\bf ii.  }{$\nabla(P^{(t)})\leftarrow\sum_{i=1}^c \lambda_i \left((H_i\oplus {\bf 0}_{n-m})^\top P^{(t)} G_{i} + (H_i\oplus {\bf 0}_{n-m}) P^{(t)} G_{i}^\top\right);$} 
\STATE{\bf iii.  }{
$Q^{(t)}\leftarrow\max_{Q\in \mathcal{D}_n}\text{trace}\left[\nabla(P^{(t)})^\top Q\right];$}
\STATE{\bf iv.  }{
$\alpha^*\leftarrow\max_{\beta\in[0,1]} \sum_{i=1}^c \lambda_i\text{trace}((H_i\oplus {\bf 0}_{n-m})Q_\alpha^{(t)}G_i(Q_\alpha^{(t)})^\top),$
where $Q_\alpha^{(t)}=\alpha P^{(t)}+(1-\alpha)Q^{(t)};$}
\STATE{\bf v.  } 
$P^{(t)}\leftarrow\alpha^* P^{(t)}+(1-\alpha^*)Q^{(t)};$
\ENDWHILE
\STATE{ 
$P^*\leftarrow\max_{P\in \Pi_n}\text{trace}\left(P^\top P^{(\text{final})}\right);$}

\STATE \textbf{Output}: $P^*$ matching multiplex graphs $\bf A$ and $\bf B$;
\end{algorithmic}
\caption{Multiplex FAQ}
\label{alg:mfaq}
\end{algorithm}

\subsection{Proof of Eq. (\ref{eq:bndprob})}
\label{app:MS}
For each $P\in \Pi_n$ define 
\begin{align}
\label{eq:xp}
X_P:&=\frac{1}{4}\left[\sum_{i=1}^c\| \widehat{A}_i -P\widehat{B}_iP^T\|^2_F-\sum_{i=1}^c\| \widehat{A}_i -\widehat{B}_i\|^2_F \right]
= \sum_{i=1}^c\frac{1}{2}\left(\text{tr}(\widehat{A}_i \widehat{B}_i)-\text{tr}(\widehat{A}_i P\widehat{B}_iP^T)\right)\\
&=\sum_{i=1}^c \sum_{\{j,\ell\}\in\Delta_P}\ha_i(j,\ell)\left[\hb_i(j,\ell)-\hb_i(\sigma_p(j),\sigma_p(\ell))\right]\notag
\end{align}
Assuming that $P\in\Pi_{n,m,k}$, then $|\Delta_P|\leq mk$.
Note that $X_P$ is a function of (at most) $3c |\Delta_P|$ independent Bernoulli random variables, and changing any one of these Bernoulli random variables can change the value of $X_P$ by at most $8$.
McDiarmid's inequality \cite{mcdiarmid1989method} then implies that for any $t\geq 0$, 
\begin{align}
\label{eq:mcd}
\p(|X_P-\Ex(X_P)|\geq t)\leq 2\text{exp}\left\{-\frac{2t^2}{192cmk}
\right\}.
\end{align}
Note that if $\widehat{C}_i(j,k), \hd_i(j,k), \hd_i(\sigma_p(j),\sigma_p(k))\in\{1,-1\}$ then
\begin{align*}
\Ex\,&\ha_i(j,\ell)\hb_i(j,\ell)=\hc_i(j,\ell)\hd_i(j,\ell)\,(1-2s_i)(1-2q_i) \\  
\Ex\,&\ha_i(j,\ell)\hb_i(\sigma_p(j),\sigma_p(\ell))=\hc_i(j,\ell)\hd_i(\sigma_p(j),\sigma_p(\ell))\,(1-2s_i)(1-2q_i). 
\end{align*}
Define
\begin{align*}
\Delta_P^{(i,0)}&=\{\,\{j,\ell\}\in\Delta_P\text{ s.t. }\hc_i(j,\ell)\neq 0  \};\\
\Delta_P^{(i,1)}&=\{\,\{j,\ell\}\in\Delta_P^{(i,0)}\text{ s.t. }0\neq\hc_i(j,\ell)\neq\hd_i(\sigma_p(j),\sigma_p(\ell))\neq 0  \};\\
\Delta_P^{(i,2)}&=\{\,\{j,\ell\}\in\Delta_P^{(i,0)}\text{ s.t. }0\neq\hc_i(j,\ell)\neq\hd_i(\sigma_p(j),\sigma_p(\ell))=0  \};\\
\Delta_P^{(i,3)}&=\{\,\{j,\ell\}\in\Delta_P^{(i,0)}\text{ s.t. }0\neq\hc_i(j,\ell)=\hd_i(\sigma_p(j),\sigma_p(\ell))  \};
\end{align*}
so that $|\Delta_P^{(i,0)}|=|\Delta_P^{(i,1)}|+|\Delta_P^{(i,2)}|+|\Delta_P^{(i,3)}|$.
We then have
\begin{align}
\label{eq:eXp}
\Ex(X_P)=&\,\sum_{i=1}^c\bigg[|\Delta_P^{(i,0)}|\,(1-2s_i)(1-2q_i)-\,|\Delta_P^{(i,3)}|\,(1-2s_i)(1-2q_i)\notag\\
&+\,|\Delta_P^{(i,1)}|\,(1-2s_i)(1-2q_i)\bigg]
\notag\\
=&\sum_{i=1}^c \left(2|\Delta_P^{(i,1)}|+|\Delta_P^{(i,2)}|\right)(1-2s_i)(1-2q_i).
\end{align}
Note that if $P,Q\in\Pi_{n,m,k}$, then $X_P=X_Q$ if $\sigma_p(j)=\sigma_q(j)$ for all $j\in[m]$; i.e., if there exists a $U\in\mathcal{P}_{m,n}$ such that $PU=Q$.
Note that this defines an equivalence relation on $P,Q\in\Pi_{n,m,k}$ which we will denote by ``$\sim$,''
and let $\Pi_{n,m,k}^*$ be a fixed (but arbitrarily chosen) set composed of one member of each equivalence class according to ``$\sim$.''
Note that $|\Pi_{n,m,k}^*|$ is at most $m^{2k}n^{2k}$.
Letting $t=\Ex(X_P)$ in Eq. (\ref{eq:mcd}), we have that if $n>\mathfrak{n}=\max(n_0,n_1)$
\begin{align*}
\p(\exists P\notin \mathcal{P}_{m,n}\text{ s.t. }X_P\leq0)&\leq
\sum_{k=1}^m\sum_{P\in \Pi_{n,m,k}^*} \p(X_P\leq 0)\\
&\leq
\sum_{k=1}^m\sum_{P\in \Pi_{n,m,k}^*} \p(|X_P-\Ex(X_p)|\geq  \Ex(X_P))\\
&\leq
\sum_{k=1}^m\sum_{P\in \Pi_{n,m,k}^*} 2\text{exp}\left\{-\frac{1344m^{1+\alpha}ck^2}{192\beta cmk}\right\}\\
&\leq
\sum_{k=1}^m\sum_{P\in \Pi_{n,m,k}^*} \text{exp}\left\{-\frac{7m^{\alpha}k}{\beta}\right\}\\
&\leq
\sum_{k=1}^m 2\text{exp}\left\{-7k\log n+2k\log n+2k\log m\right\}\\
&\leq 2\text{exp}\left\{-2\log n\right\}
\end{align*}
as desired.


\subsection{Proof of Eq. (\ref{eq:mcd3})}
We have that if each $W_i\sim$ER$(n,p_i)$, then 
$$\Delta_P^{(i,1)}=\sum_{\{j,\ell\}\in\Delta_P}\mathds{1}\{\hc_i(j,\ell)\neq \hd_i(\sigma_p(j),\sigma_p(\ell))\},$$
so
that $\Ex(\Delta_P^{(i,1)})=2p_i(1-p_i)|\Delta_P|$.
Also, $\Delta_P^{(i,1)}$ is then a function of at most $2|\Delta_P|$ independent Bernoulli random variables, and changing the value of any one these can change the value of $\Delta_P^{(i,1)}$ by at most $2$.
McDiarmid's inequality then yields the desired result 
\begin{align}
\label{eq:mcd2}
\p(|\Delta_P^{(i,1)}-\Ex(\Delta_P^{(i,1)})|\geq t)\leq 2\text{exp}\left\{-\frac{2t^2}{8|\Delta_P|}
\right\},
\end{align}
by setting $t= p_i(1-p_i)|\Delta_P|.$


\subsection{Proof details for Eq. \ref{eq:erms}}
\label{app:erms}

Let the equivalence relation ``$\sim$'' on  $\Pi_{n,m,k}$ be defined via $P\sim Q$ if there exists a $U\in\mathcal{P}_{n,m}$ such that $PU=Q$.
Note that if $P\sim Q$ then 
$$\sum_{i=1}^{c}\|(\widehat{A}_i\oplus \textbf{0}_{n-m})P-P\widehat{B}_i\|^2_{F}=\sum_{i=1}^{c}\|(\widehat{A}_i\oplus \textbf{0}_{n-m})Q-Q\widehat{B}_i\|^2_{F}.$$
Let $\Pi_{n,m,k}^*$ be a fixed (but arbitrarily chosen) set composed of one member of each equivalence class according to ``$\sim$,'' and 
note that $|\Pi_{n,m,k}^*|$ is at most $m^{2k}n^{2k}$.
Given the assumptions in Section \ref{sec:MSmatch}, for $n>n_2$ we have that for
each $P\in\Pi_{n,m,k}^*$,
\begin{align}
\label{eq:deltaPkER}
\p&\bigg(\bigcup_{i=1}^c \left\{|\Delta_P^{(i,1)}|<\frac{1}{3}mkp_i(1-p_i)\right\}\bigg)\notag
\\&\leq \sum_{i=1}^c  \,2\text{exp}\left\{\frac{-2 mkp_i^2 }{96}\right\}\notag\\
&\leq 2\text{exp}\left\{-8 k\log n +\log c\right\}\leq 2\text{exp}\left\{-7 k\log n\right\}.
\end{align}
Denote the event bound in Eq. (\ref{eq:deltaPkER}) via $\mathcal{E}_{n,P}$.

For $n>\max(n_2,n_0)$, we then have
\begin{align*}
\p&\bigg(\argmin_{P\in\Pi_{n}}\sum_{i=1}^{c}\|(\widehat{A}_i\oplus \textbf{0}_{n-m})P-P\widehat{B}_i\|^2_{F}\not\subset \mathcal{P}_{m,n} \bigg)=\p(\exists P\notin \mathcal{P}_{m,n}\text{ s.t. }X_P\leq0)\\
&\leq\sum_{k=1}^m\sum_{P\in \Pi_{n,m,k}^*} \p(X_P\leq 0)\\
&\leq\sum_{k=1}^m\sum_{P\in \Pi_{n,m,k}^*} \p(|X_P-\Ex(X_p)|\geq  \Ex(X_P)\cap \mathcal{E}_{n,P}^c)+\p(\mathcal{E}_{n,P})\\
&\leq
\sum_{k=1}^m\sum_{P\in \Pi_{n,m,k}^*} 2\text{exp}\left\{-\frac{1344m^{1+\alpha}ck^2}{192\beta cmk}\right\}+2\text{exp}(-7k\log n)\\
&\leq\sum_{k=1}^m4\text{exp}(-7k\log n+2k\log n+2k\log m)\leq 4n^{-2},\\
\end{align*}
as desired.


\subsection{Proof details for Eq. (\ref{eq:probME}) }
\label{app:pfME}

For $P\in\Pi_n$, define
\begin{align*}
\Delta_P^{(1)}&:=\{ \{j,\ell\}\in\Delta_P\text{ s.t. }T(j,\ell)=1; W(\sigma_p(j),\sigma_p(\ell))=0  \};\\
\Delta_P^{(2)}&:=\{ \{j,\ell\}\in\Delta_P\text{ s.t. }T(j,\ell)=0; W(\sigma_p(j),\sigma_p(\ell))=1  \};\\
\Delta_P^{(3)}&:=\{ \{j,\ell\}\in\Delta_P\text{ s.t. }T(j,\ell)=W(\sigma_p(j),\sigma_p(\ell))=1  \};\\
\Delta_P^{(4)}&:=\{ \{j,\ell\}\in\Delta_P\text{ s.t. }T(j,\ell)=W(\sigma_p(j),\sigma_p(\ell))=0  \};\\
e_P&:=\{ \{j,\ell\}\in\Delta_P\text{ s.t. }T(j,\ell)=1 \};\\
\mathfrak{n}_P&:=\{ \{j,\ell\}\in\Delta_P\text{ s.t. }T(j,\ell)=0  \}.
\end{align*}
For $P\in\Pi_{n,m,k}$, we then have that $X_P$ defined in Eq. (\ref{eq:xp}) satisfies
\begin{align*}
\Ex(X_P)=&(|e_P|-|\Delta_P^{(3)}|)\sum_i (1-2s_i)(1-2r_i)+(|\mathfrak{n}_P|-|\Delta_P^{(4)}|)\sum_i(1-2q_i)(1-2t_i)\\
&+|\Delta_P^{(1)}|\sum_i (1-2s_i)(1-2t_i)+|\Delta_P^{(2)}|\sum_i (1-2q_i)(1-2r_i)\\
=& |\Delta_P^{(1)}|\sum_i 2(1-2s_i)(1-r_i-t_i)+|\Delta_P^{(2)}|\sum_i 2(1-2q_i)(1-r_i-t_i).
\end{align*}

\end{document}